\documentclass[
 reprint,
 superscriptaddress,
 amsmath,amssymb,
 aps,
 prl
]{revtex4-2}

\usepackage{graphicx}% Include figure files
\usepackage{dcolumn}% Align table columns on decimal point
\usepackage{bm}% bold math
\usepackage[hidelinks,colorlinks=true,linkcolor=blue,citecolor=blue, anchorcolor = blue, urlcolor = blue]{hyperref}
\usepackage[mathlines]{lineno}% Enable numbering of text and display math

\usepackage{braket}
\usepackage{color}
\usepackage{float}

\newcommand{\affsiqse}{\affiliation{Shenzhen Institute for Quantum Science and Engineering (SIQSE), Southern University of Science and Technology, Shenzhen, P. R. China.}}
\newcommand{\affiqa}{\affiliation{International Quantum Academy, Shenzhen 518048, China.}}
\newcommand{\affgdlab}{\affiliation{Guangdong Provincial Key Laboratory of Quantum Science and Engineering, Southern University of Science and Technology Shenzhen, 518055, China.}}

\newcommand{\yb}{^{171} \mathrm{Yb}^+}
\newcommand{\figref}[1]{Fig.~\ref{#1}}
\newcommand{\equref}[1]{Eq.~(\ref{#1})}

\newcommand{\sket}[1]{\ket{^{2}\mathrm{S}^#1_{1/2}}}
\newcommand{\sbra}[1]{\bra{^{2}\mathrm{S}^#1_{1/2}}}
\newcommand{\pket}[1]{\ket{^{2}\mathrm{P}^#1_{1/2}}}
\newcommand{\pbra}[1]{\bra{^{2}\mathrm{P}^#1_{1/2}}}

\definecolor{revise}{RGB}{112,48,160}

\begin{document}

\title{Observing Quantum Synchronization of a Single Trapped-Ion Qubit}

\affsiqse
\affiqa
\affgdlab

\author{Liyun Zhang}
 \thanks{These two authors contributed equally to this work.}

\author{Zhao Wang}
 \thanks{These two authors contributed equally to this work.}

\author{Yucheng Wang}

\author{Junhua Zhang}

\author{Zhigang Wu}

\author{Jianwen Jie}
 \email{Jianwen.Jie1990@gmail.com}

\author{Yao Lu}
 \email{luy7@sustech.edu.cn}

\date{\today}

\begin{abstract}
Synchronizing a few-level quantum system is of fundamental importance to the understanding of synchronization in deep quantum regime. Whether a two-level system, the smallest quantum system, can be synchronized has been theoretically debated for the past several years. Here, for the first time, we demonstrate that a qubit can indeed be synchronized to an external driving signal by using a trapped-ion system. By engineering fully controllable gain and damping processes, an ion qubit is locked to the driving signal and oscillates in phase. Moreover, upon tuning the parameters of the driving signal, we observe characteristic features of the Arnold tongue as well. Our measurements agree remarkably well with numerical simulations based on recent theory on qubit synchronization. By synchronizing the basic unit of quantum information, our research unlocks potential applications of quantum synchronization in quantum information processing.
\end{abstract}

\maketitle

\textit{Introduction.---}
Originally discovered by Huygens in two pendulum clocks suspended from the same wooden beam \cite{bell1941horologium}, synchronization is widespread in nature \cite{PhysRevLett.64.821,EC1996,schafer1998heartbeat,PhysRevLett.103.168103,fell2011role,PhysRevLett.112.014101}. Systems capable of synchronization usually suffer from certain dissipations, holding a common characteristic known as the \textit{limit cycle}. It represents the steady motion of the system at the intrinsic oscillating frequency, possessing one or more free phases and being robust to perturbations. Owning a limit cycle enables a self-sustained oscillator to adjust its rhythm to oscillate in phase via mutual coupling to other oscillators or being driven by an external signal \cite{pikovsky2001synchronization, andronov2013theory}.

As controllable quantum systems blossom in recent decades \cite{ladd2010quantum,preskill2018quantum,PRXQuantum.2.017001}, synchronization studies have also been extended to the quantum regime \cite{zhirov2006quantum,PhysRevLett.97.210601, PhysRevLett.100.014101, lohe2010quantum, PRL2013OM, PRL2013Vdp, PRL2014Vdp, PhysRevLett.113.154101, eneriz2019degree}. Several works focus on the quantum van der Pol (qvdP) oscillator \cite{PRL2013Vdp,PRL2014Vdp,PRE2015Vdp,Adp2015Vdp,PRL2017kerr,PRL2018VdP,PRL2019VdP,PRE2020Vdp,PRR2020VdP,PRR2021Vdp}. Compared to its classical counterpart \cite{Vdp1926}, the deterministic trajectory and the limit cycle becomes meaningless due to the quantum noise \cite{PRL2013Vdp, PRL2014Vdp,PRE2020Vdp}. Nevertheless, the synchronization features can still be captured by evaluating the phase preference of the quasi-probability distribution in the position-momentum space \cite{PRL2013Vdp, PRL2014Vdp, PRL2019VdP, PRE2020Vdp}.
Meanwhile, inspired by the qvdP oscillator in the deep quantum regime where only a few discrete energy levels are involved \cite{PRL2013Vdp,PRL2014Vdp,PRE2015Vdp,PRL2019VdP}, the synchronization of systems with finite Hilbert space has also been of interest.
Raised by Roulet and Bruder, spin-1 systems were theoretically shown to be synchronizable \cite{PRL2018, PRL2018QN,PRA2019tribit}, and the experimental demonstrations were carried out subsequently \cite{PRL2020exp,PRR2020SPin1}.
Here, the limit cycle and synchronization can be generalized to spin systems by introducing the spin coherent states and Husimi-\textit{Q} representation \cite{PRL2018, PhysRevA.12.1019}. The emergence of coherence between different energy levels induced by the external signal can be also interpreted as a sign of phase synchronization \cite{PRA2019tribit}, providing a fresh understanding of quantum synchronization.

\begin{figure}[htbp!]
\centering
\includegraphics[scale = 1]{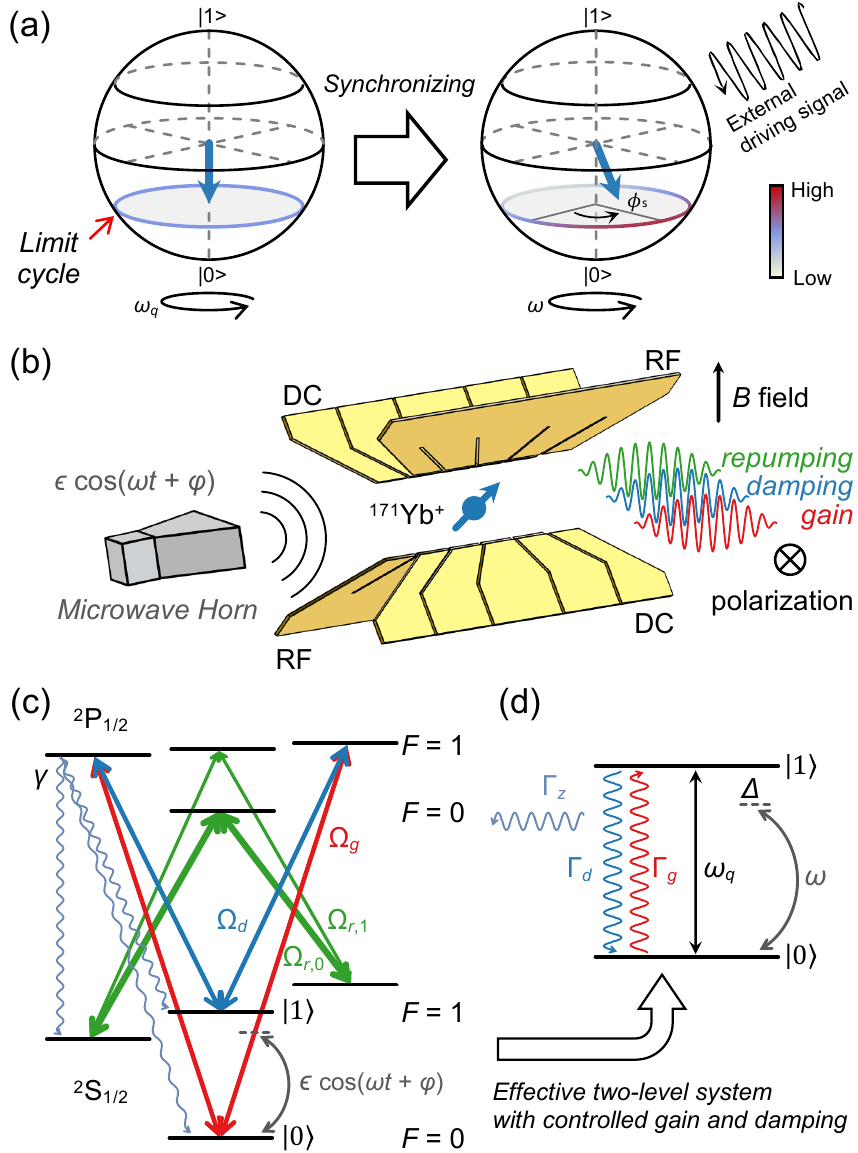}
\caption{\label{fig:fig1}
(a). Visualizing limit cycle and synchronization on the Bloch sphere.
A limit cycle can be constructed by mixing pure states in the latitude circle with equal weights (uniform blue circle in the left sphere). All these states precess at the intrinsic frequency of $\omega_q$. After synchronization, the qubit obtains a phase preference $\phi_\mathrm{s}$ and precesses at the sync frequency $\omega$. The blue arrows indicate the Bloch vectors of the corresponding qubit states.
(b). Experimental setup. A single $\yb$ ion is held in a 5-segmented blade-type Paul trap. Three beams, marked as \textit{gain}, \textit{damping} and \textit{repumping} are aligned to the ion, and all the polarizations are perpendicular to the static magnetic field. The external sync signal is broadcasted to the ion through a microwave horn.
(c). Energy levels of $\yb$ ion. The transitions driven by the \textit{gain} (red), \textit{damping} (blue) and \textit{repumping} (green) beams are shown, with the coupling strengths of $\Omega_g$, $\Omega_d$ and $\Omega_{r,0/1}$ respectively. For the spontaneous emission we only show the decay channels from the $^2P_{1/2}\ket{F = 1, m_F = - 1}$ level for clarity.
(d). Dissipated qubit. Provided that $\gamma \sim \Omega_{r,0} \gg \Omega_g, \Omega_d,\Omega_{r,1}$, as is the case in our experiment, the whole system is well approximated by a two-level system with controlled gain and damping rates, which can be coherently driven by the sync signal.}
\end{figure}

However, Ref. \cite{PRL2018} also stated that quantum synchronization is not applicable for a single qubit (spin-1/2 system) due to the lack of the limit cycle \cite{kwek2018no}. 
It contradicts the fact that the qvdP oscillator can be synchronized at the quantum limit, where only the ground and the first-excited states are occupied \cite{PRL2013Vdp, PRL2014Vdp, PRE2015Vdp}. 
Parra-L\'{o}pez and Bergli solved this seeming conflict theoretically, showing that the limit cycle of a single qubit can be obtained by choosing appropriate pure states to construct mixed states [see \figref{fig:fig1}~(a)] \cite{PRA2020two}; thus, a single qubit is certainly synchronizable. 
The debate on whether qubits can be synchronized is not only of academic interest; being able to synchronize qubits holds promise for potential applications of quantum synchronization in quantum information \cite{lohe2010quantum,PhysRevA.94.052121,buca2022algebraic}.

In this work, we experimentally demonstrate that a qubit can be synchronized with an external driving signal by utilizing a trapped-ion system.
We provide a clear characterization of the limit cycle, supply strong evidence of synchronization and ascertain the parameter intervals within which synchronization can be achieved. Our results are in remarkable quantitative agreement with theoretical predictions, showcasing trapped-ion systems as an excellent platform for further studies and applications of quantum synchronization.

\textit{Experimental realization.---}
We demonstrate the quantum synchronization in a qubit encoded in a single $\yb$ ion [see \figref{fig:fig1}~(b)]. Here, the qubit is represented by the hyperfine levels belonging to the ground manifold, denoted as $\ket{0} \equiv {}^2S_{1/2}\ket{F = 0, m_F = 0}$ and $\ket{1} \equiv {}^2S_{1/2}\ket{F = 1, m_F = 0}$, with an energy gap of $\omega_q / (2\pi)$ around $12.6$~GHz \cite{fisk1997accurate}, as shown in \figref{fig:fig1}~(c). The qubit can be initialized to the $\ket{0}$ state via optical pumping, while the state measurement is performed by selectively exciting the $\ket{0}$ level to the $^2P_{1/2}\ket{F = 0}$ levels and counting the emissing photons \cite{PRA2007Yb}. In our system, the state-preparation-and-measurement (SPAM) error is less than $7\times10^{-3}$. More details of the setup can be found in the Supplemental Material (S.M.)~I. \cite{suppmaterials}

To realize a dissipated qubit, we engineer the gain and damping processes in a fully controlled way \cite{li2019observation, PhysRevLett.126.083604}, as illustrated in \figref{fig:fig1}~(c) and (d). More specifically, the ion is driven from $\ket{0}$ ($\ket{1}$)  to  $^2P_{1/2}\ket{F = 1, m_F = \pm 1}$ at a coupling strength of $\Omega_g$ ($\Omega_d$), followed by a fast spontaneous emission to both $\ket{1}$ and $\ket{0}$. The state leakage to $^2S_{1/2}\ket{F = 1, m_F = \pm 1}$ is quickly repumped back to the qubit space by strongly coupling these states to $^2P_{1/2}\ket{F = 0, m_F = 0}$ and $^2P_{1/2}\ket{F = 0, m_F = 1}$. Taken together, these processes result in an incoherent gain (damping) at a rate of $\Gamma_g$ ($\Gamma_d$) within the qubit system, accompanied by a pure dephasing dynamics at a rate of $\Gamma_z$.

The dissipation rates of $\Gamma_g$ and $\Gamma_d$ can be independently tuned by adjusting the related lasers' power to change strengths of $\Omega_g$ and $\Omega_d$, respectively, and the dephasing rate of $\Gamma_z$ is determined afterwards. In the experiments, these rates are set by choosing appropriate coupling strengths and accurately measured. Further details can be found in the S.M.~I and II \cite{suppmaterials}.

A classical external signal (referred to the sync signal in following) with frequency $\omega$, strength $\epsilon$ and initial phase $\varphi = \pi/2$ is then broadcasted to the ion through a microwave horn \cite{PRA2007Yb}.
In the rotating frame of $\hat{U}_\mathrm{sync} = e^{-i\omega \hat{\sigma}_z t /2}$, the dynamics of the qubit can be described by the following master equation ($\hbar = 1$) \cite{PRA2012effective, suppmaterials},
\begin{eqnarray}\label{equ:qsstate}
 \frac{d\hat{\rho}}{dt}=- \dfrac{i}{2} \left[ \Delta \hat{\sigma}_z + \epsilon \hat{\sigma}_y, \hat{\rho} \right]+ \mathcal{L}_{0}\hat{\rho},
\end{eqnarray}
where $\hat \rho$ is the density operator, $\Delta = \omega_q - \omega$ is the frequency detuning,  $\hat \sigma_{x,y,z}$ are Pauli operators and $\mathcal{L}_{0}\hat{\rho}=(\Gamma_g \mathcal{D}[\hat{\sigma}_+] + \Gamma_d \mathcal{D}[\hat{\sigma}_-] + \Gamma_{z} \mathcal{D}[\hat{\sigma}_z])\hat{\rho}/2$
describes the effective dissipations with Lindblad super-operator $\mathcal{D}[\mathcal{\hat{A}}]\hat{\rho} = \mathcal{\hat{A}}\hat{\rho} \mathcal{\hat{A}}^\dagger - \{ \mathcal{\hat{A}}^\dagger \mathcal{\hat{A}}, \hat{\rho} \}/2$.
The signatures of the synchronization are evaluated in the rotating frame to avoid directly measuring the fast oscillation of the qubit in the lab frame [see in S.M.~IV \cite{suppmaterials}], and experimental results are compared to numerical simulations based on Eq.~(\ref{equ:qsstate}).

\begin{figure*}[htbp]
\centering
\includegraphics[scale = 1.]{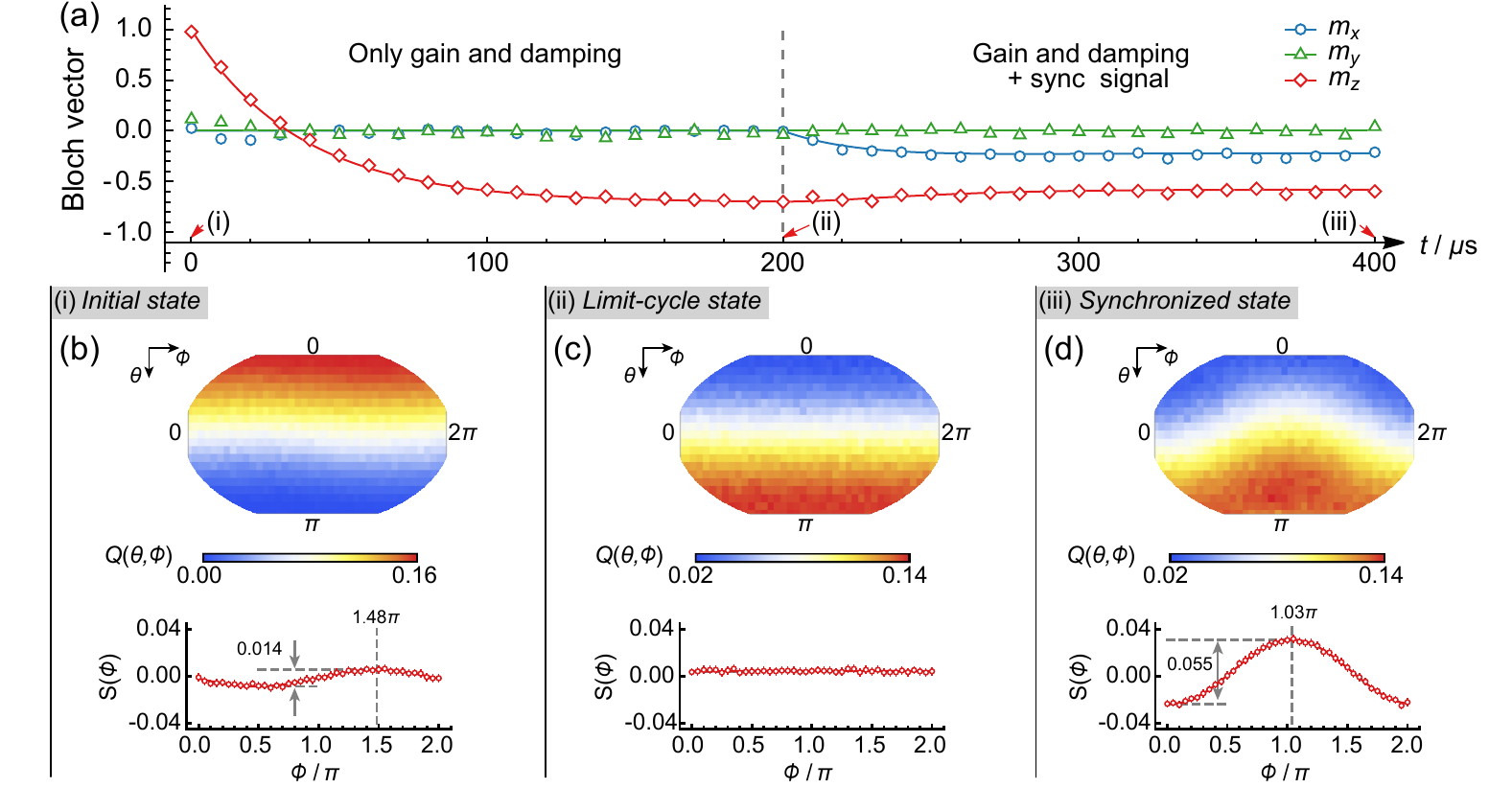}
\caption{\label{fig:fig2} Experimental results of quantum synchronization.
(a). Time evolution of the Bloch vector. From $0~\mathrm{\mu s}$ to $200~\mathrm{\mu s}$, only gain and damping processes are engineered to establish the limit cycle. Subsequently from $200~\mathrm{\mu s}$ to $400~\mathrm{\mu s}$, the sync signal is triggered simultaneously to synchronize the qubit to the signal. Here and in the following figures, the open markers represent the experimental results, while the error bars of a standard deviation are smaller than the size of the markers. The solid lines represent numerical simulation results. To verify the achievements of the limit cycle and the synchronized state, we measure the \textit{Q}-functions and also the \textit{S}-functions at the time of (i) $t_0=0~\mathrm{\mu s}$, (ii) $t_1=200~\mathrm{\mu s}$ and (iii) $t_1 + t_2=400~\mathrm{\mu s}$, which correspond to the initial, limit cycle and synchronized state, respectively. (b)-(d) shows the results of the \textit{Q}-function under the Winkel tripel projection (first row) and the \textit{S}-function (second row) for the time of (i)-(iii), respectively.}
\end{figure*}

\textit{Results.---}
The experimental observations of the qubit synchronization are summarized in \figref{fig:fig2}. We prepare the qubit to the $\ket{1}$ state and confirm that the undriven self-sustained qubit would relax to the limit cycle state under the dissipations. Here for demonstration, the values of $\Gamma_g$, $\Gamma_d$ and $\Gamma_z$ are set to be $(2\pi)\times1.27(4)$~kHz, $(2\pi)\times7.33(11)$~kHz and $(2\pi) \times 4.42(106)$~kHz respectively, giving an anisotropic ratio $\Gamma_d / \Gamma_g = 5.8$. The measurements of these rates are discussed in S.M.~III~\cite{suppmaterials}, and the error bars here and below all represent standard deviation. Note that the synchronization can occur as long as $\Gamma_g \neq \Gamma_d$ \cite{PRA2020two}.

As shown in \figref{fig:fig2}~(a), after the first-stage evolution lasting $200~\mathrm{\mu s}$, the qubit has reached the limit cycle described by the Bloch vector of $\mathbf{m}_\mathrm{LC}^{\mathrm{(exp)}} = \{ 0, 0, -0.700(16) \}$,
which is consistent with the predicted value of $\mathbf{m}_\mathrm{LC}^{\mathrm{(th)}} =\{ 0, 0, ({\Gamma_g - \Gamma_d})/{(\Gamma_g + \Gamma_d)} \} =  \{ 0, 0, -0.705 \}$.
The limit cycle can be further visualized by utilizing the Husimi-\textit{Q} representation \cite{PhysRevA.12.1019}, and the related \textit{Q}-function reads as below,
\begin{equation}\label{equ:qfunc}
  Q(\theta,\phi) = \dfrac{1}{2\pi} \bra{\theta,\phi}\hat\rho\ket{\theta,\phi}.
\end{equation}
Here $\ket{\theta,\phi}$ is the coherent spin state, while for the qubit, it corresponds to the eigenstate of the operator $\mathbf{n}_{\theta, \phi} \cdot \hat{\bm{\sigma}}$ and $\mathbf{n}_{\theta, \phi} = \{ \sin\theta\cos\phi, \sin\theta\sin\phi, \cos\theta \}$.
Similar to the Wigner function for describing harmonic oscillator, the \textit{Q}-function represents the quasi-probability distribution of any qubit state in the phase portrait of the Bloch sphere. The \textit{Q}-functions of the initial and limit cycle states are experimentally measured, as illustrated in \figref{fig:fig2}~(b) and (c). The contribution to the limit cycle mainly comes from the pure states near the south pole of the Bloch sphere, due to the anisotropic ratio $\Gamma_d / \Gamma_g > 1$. The preference in the phase $\phi$ of any state can be evaluated by the \textit{synchronization measurement}, denoted as the \textit{S}-function \cite{PRL2018, PRA2020two},
\begin{equation}\label{equ:sfunc}
  S(\phi) = \int_{0}^{\pi} Q(\theta, \phi) \sin\theta \mathrm{d}\theta - \dfrac{1}{2\pi}.
\end{equation}
The values of $S$-function for the initial and limit cycle states are shown in \figref{fig:fig2}~(e) and (f). The initial state shows a weak phase preference around $1.48(2)\pi$ due to the imperfect state preparation, while for the limit cycle this preference completely vanishes as it should.

After the qubit reaches at the limit cycle, we apply a resonant sync signal ($\Delta \sim 0$) with strength $\epsilon = (2\pi)\times 2.37(1)$~kHz ($\epsilon/\Gamma_g=1.87$). In \figref{fig:fig2}~(a), the qubit has reached the final synchronized state after another $200~\mathrm{\mu s}$ evolution. As a sign of synchronization, a non-zero value is observed in $m_x$, indicating that the sync signal rebuilds the qubit coherence. For further verification, we also measure the \textit{Q}- and \textit{S}-functions for the synchronized state, as shown in \figref{fig:fig2}~(d) and (g). The phase preference reappears at $\phi = 1.03(2)\pi$, since we set $\Gamma_d > \Gamma_g$ and the sync signal phase to be $\pi/2$ [see in S.M.~IV \cite{suppmaterials}]. The fitting contrast of the $S$-function reaches 0.055(1), significantly deviating from the unsynchronized limit cycle. The above results indicate the experimental achievement of synchronization with anti-phase locking.

\begin{figure}[htbp]
\centering
\includegraphics[scale = 1.]{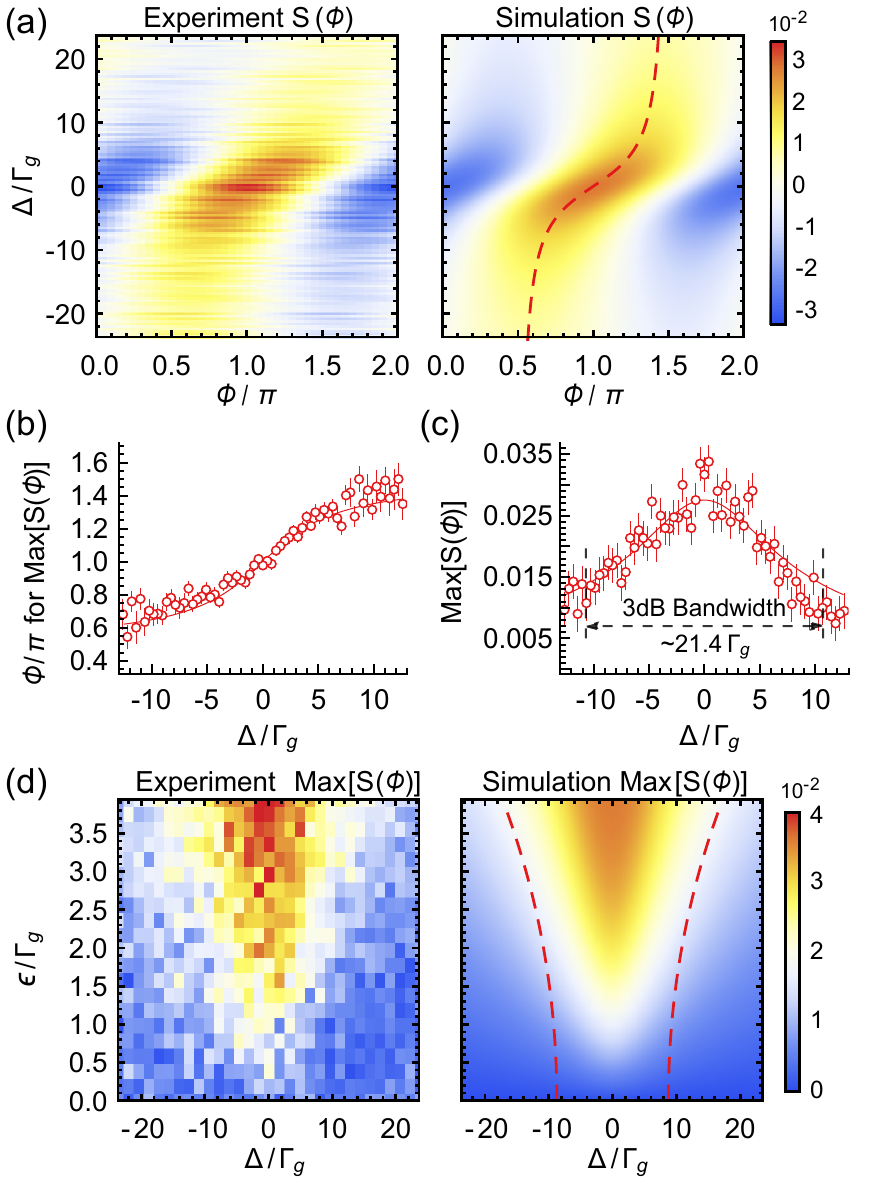}
\caption{\label{fig:fig3}
(a). Synchronization under detuned sync signals. Here the sync strength is kept at $\epsilon=1.87\Gamma_{g}$. The experimental results (left) agree well with simulations (right). The synchronization performs the best under the resonant situation and can be achieved across a wide range of sync frequency.
(b). Shift in locked phase. The numerical simulation results correspond to those along the red-dashed in the right sub-figure of (a).
(c). Frequency bandwidth of synchronization. A 3dB bandwidth of $21.4\Gamma_g$ is obtained at the sync strength of $\epsilon=1.87\Gamma_{g}$.
(d). Arnold tongue. The left and right sub-figures correspond to the experimental and the simulation results. The red dashed lines in the right sub-figure indicate the 3dB frequency bandwidth of the synchronization under different sync strengths.}
\end{figure}

Now we tune the sync frequency away from the qubit frequency ($\Delta \neq 0$) while leaving the sync strength $\epsilon$ unchanged to investigate the frequency entrainment. \figref{fig:fig3}~(a) clearly shows that the qubit can be synchronized for a wide range of the sync signal frequency but with a delay or advance in the locked phase [see \figref{fig:fig3}~(b)]. 
Such phase shift might provide an alternative approach to sense the drift in magnetic field strength by employing a magnetic-field-sensitive qubit.

In addition to the locked phase shift, the maximal value of the synchronization measurement, $\mathrm{Max}[S(\phi)]$, also degrades with increasing detuning $\Delta$. As an example shown in \figref{fig:fig3}~(c), the value of $\mathrm{Max}[S(\phi)]$ drops by half when the detuning $\Delta$ increases to $10.7\Gamma_g$, under the sync strength of $\epsilon=1.87\Gamma_g$. This frequency bandwidth strongly depends on the sync strength; therefore, we measure the value of $\mathrm{Max}[S(\phi)]$ across a wide range of the sync detuning and strength, as shown in \figref{fig:fig3}~(d). This result is known as the \textit{Arnold tongue} \cite{pikovsky2001synchronization}. It is remarkable that synchronization occurs as long as the sync strength is non-zero, regardless of how weak it is. However, weak signals result in small maximal values in the \textit{S}-function, making the synchronization hard to be distinguished from the noisy fluctuation in practice.

\begin{figure}[tbp]
  \centering
  \includegraphics[scale = 1.]{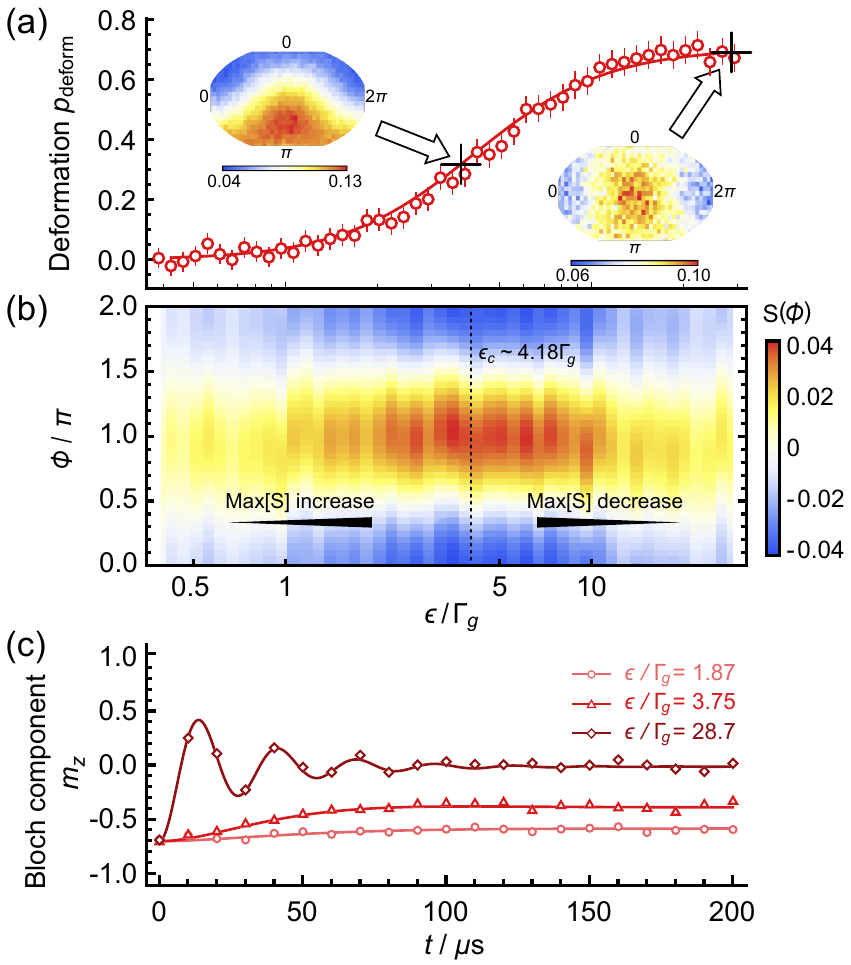}
  \caption{\label{fig:fig4} Performance of synchronization under resonant driving with different sync strengths. (a). Deformation of limit cycle. The synchronized state departs from the limit-cycle state as the sync strength increases. The left and right inset figures indicate the \textit{Q}-functions with the sync strength $\epsilon$ of $3.75\Gamma_g$ and $28.7\Gamma_g$, respectively. (b). Synchronization measurement under different sync strengths. The maximal value of the \textit{S}-function increases at first, and then after a critical value of $\epsilon_{c}$ it tends to drop back to zero. Under our setting parameters, the critical sync amplitude is around $4.18\Gamma_g$. (c). Time evolution of Bloch vector component $m_z$ starting from the limit cycle state. Under the sync strengthes of $1.87\Gamma_g$ and $3.75\Gamma_g$, the system evolves almost smoothly to the final steady state, while with the strength of $28.7\Gamma_g$, the system significantly oscillates before fully decay.}
\end{figure}

The limit cycle can also be distorted by sync signals that are too strong. The deformation of the limit cycle can be characterized by \cite{PRA2019tribit},
\begin{equation}\label{equ:deform}
  p_\mathrm{deform}(\epsilon) = \mathrm{Tr}\left[ \hat{\sigma}_z ( \hat{\rho}_\mathrm{syn}(\epsilon) - \hat{\rho}_\mathrm{syn}(0) ) \right],
\end{equation}
where $\hat{\rho}_\mathrm{syn}(\epsilon)$ is the density matrix of the synchronized state under the sync strength of $\epsilon$, and $\hat{\rho}_\mathrm{syn}(0)$ indicates the limit cycle.
We experimentally measure the deformation with the resonant sync signal, shown in \figref{fig:fig4}~(a). The deformation increases as the sync strength gets stronger, and eventually saturates to $p_\mathrm{deform}^\mathrm{(sat)} = - m_{\mathrm{LC},z}$. The saturation indicates that the Bloch vector of the qubit state lies on the x-y plane of the Bloch sphere, and the limit cycle established by the gain and damping is fully distorted. This matches the results of the corresponding \textit{S}-function measurement in \figref{fig:fig4}~(b). We find that the phase preference becomes first more pronounced as the sync strength increases and then gradually weakens when the signal becomes too strong.
Thus there is generally a trade-off between the distortion in the limit cycle and the sensitivity of the phase preference, which should be carefully balanced in practical implementations.

We have further checked the time evolution of the qubit starting from the limit cycle state but under different sync strengths, as illustrated in \figref{fig:fig4}~(c). The significant oscillation of the system energy is observed when the sync strength approaches $28.7\Gamma_g$, and the fitting result gives the oscillating frequency of $27.7(6)\Gamma_g$. This result clearly indicates that the qubit is forcibly driven by the too strong signal, instead of synchronizing to it.

\textit{Conclusion and discussion.---}
We have realized the synchronization between an ion qubit and an external driving signal, and clear evidence of phase synchronization and frequency entrainment has been observed.
The methods developed here can be extended to various platforms, such as neutral atoms and superconducting qubits, making it possible to synchronize hybrid systems with proper frequency conversion.

Future research should further consider synchronization in multi-qubits systems, as well as finding practical applications.
For instance, qubits synchronized to the same signal can be treated as a homogeneous spin ensemble, and such ensembles in quantum magnetc could be helpful to generate phase-correlated magnetic fields and enhance quantum sensing \cite{buca2022algebraic}. 
For quantum memories in solid-state systems, the memory qubits inevitably couple to the environmental qubits with random phases and strengths \cite{zhong2015optically}. Synchronizing environmental qubits might effectively suppress the randomness imparted to memory qubits, thereby prolonging their coherent storage time.
Recent research also widely considers utilizing dissipations as resources for quantum information \cite{verstraete2009quantum,Harrington2022wbi}, and quantum synchronization offers one fresh and promising route for applications of dissipative quantum engineering.
\\

\begin{acknowledgments}
  We gratefully thank Dapeng Yu for providing critical support to this project. 
  We also thank Junqiu Liu for carefully reading this manuscript and Fudong Wang for the helpful discussions on quantum memory. 
  
  This work was supported by the National Science Foundation of China (Grants No.~12004165, No.~12104210), 
  Shenzhen Science and Technology Program (Grants No.~RCBS20200714114820298), 
  Guangdong Basic and Applied Basic Research Foundation (Grant No. 2022B1515120021),
  China Postdoctoral Science Foundation (Grant No.~2022M711496),  
  Guangdong Provincial Key Laboratory (Grant No.~2019B121203002).
  Z. Wu ackonwleges support from the National Science Foundation of China (No.~11974161) and Shenzhen Science and Technology Program (Grants No.~KQTD20200820113010023)
  Y. Wang ackonwleges support from the National Science Foundation of China (No.~12104205).
\end{acknowledgments}

\bibliography{QS_Refs}

\setcounter{equation}{0} \setcounter{figure}{0}
\setcounter{table}{0} %\setcounter{page}{1} \makeatletter
\renewcommand{\theparagraph}{\bf}
\renewcommand{\thefigure}{S\arabic{figure}}
\renewcommand{\theequation}{S\arabic{equation}}

\onecolumngrid
\flushbottom
%%%%%%%%%%%%%%%%%%%%%%%%%%%%%%%%%%%%%%%%%%%%%%%%
\newpage

\section*{Supplemental Material}

\section{Theory}

\section{Experimental setup}

\begin{figure}[htbp]
  \centering
  \includegraphics[scale = 0.9]{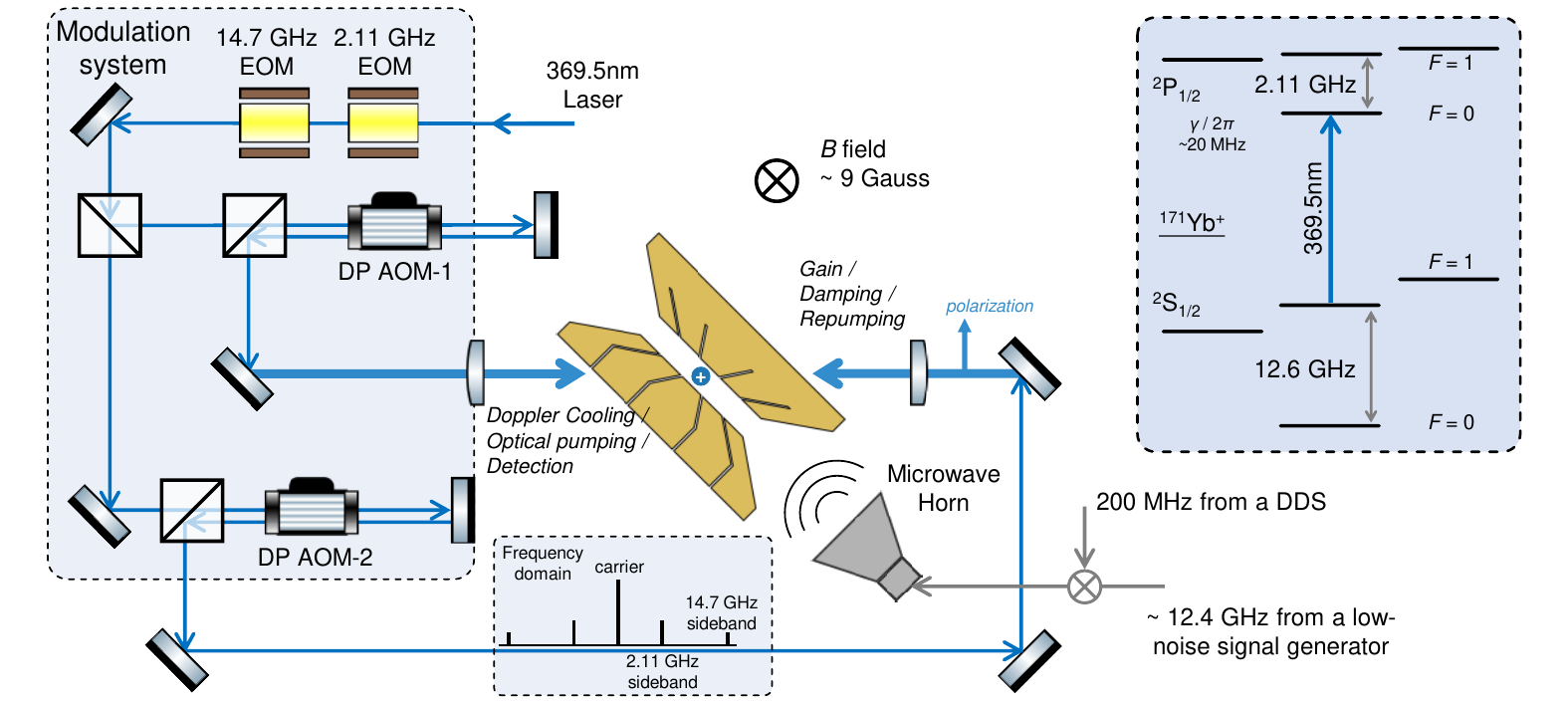}
  \caption{\label{fig:Suppfig1} Experimental setup. A single ion is trapped in the blade-type Paul trap. A 369.5~nm laser beam, which drives the main transition between the $^2S_{1/2}$ and $^2P_{1/2}$ levels (as shown in the upper-right inset figure), goes through a modulation system and then aligned to the ion position, to realize all the operations required in the experiments. A static magnetic field \textit{B} around 9~Gauss is set to define the quantization axis and lift the energy degeneracy of the Zeeman levels as well. The external driving signal, which is broadcasted to the ion through the microwave horn, is generated by mixing a fixed microwave signal (around 12.4~GHz) from a ultra low-noise signal generator with a tunable radiofrequency signal (around 200~MHz) from a direct digital synthesis board (DDS).}
\end{figure}

In the main text, we have briefly demonstrated the experimental system. Here, we add more detials of the setup, as shown in \figref{fig:Suppfig1}. We capture a single Ytterbium ion with trap frequencies of approximately $\{\nu_x, \nu_y, \nu_z\} = 2\pi \times \{ 2.8, 2.2, 0.5 \}~\mathrm{MHz} $. For $\yb$, the resonant wavelength between its ground state $^2S_{1/2}$ and the first excited state $^2P_{1/2}$ is about 369.5~nm. To cover all the hyperfine structues of the $\yb$, the 369.5~nm laser is delivered through a modulation system consisting of two electro-optical modulators (EOMs) and two acoustic-optical modulators (AOMs). The two EOMs phase modulate the passing laser at 2.11~GHz and 14.7~GHz, respectively; thus, the laser will acquire sidebands in the frequency domain corresponding to the modulation frequencies after passing through them. When the non-modulated carrier frequency is tuned to resonant with the $^2S_{1/2}\ket{F = 1} \leftrightarrow  {^2}P_{1/2}\ket{F = 0}$ transition, the 1st order sideband from the 2.11~GHz EOM is resonant with the $^2S_{1/2}\ket{F = 1} \leftrightarrow {^2}P_{1/2}\ket{F = 1}$ transition, and that from the 14.7~GHz (2.11~GHz + 12.6~GHz) EOM is resonant with the transition of $^2S_{1/2}\ket{F = 0} \leftrightarrow {^2}P_{1/2}\ket{F = 1}$. The two AOMs are used here as fast optical shutters and frequency shifters, set up in a double-pass (DP) configuration. The switching of all the signal sources used for the EOMs and AOMs is controlled by TTL signals, allowing us to quickly turn on/off each device at the nanosecond level, thus applying a series of different operations accurately in the time sequence.

The laser beam diffracted from the AOM-1 is used to implement the processes of Doppler cooling (14.7~GHz EOM and AOM-1 on simultaneously), optical pumping (2.11~GHz EOM and AOM-1 on simultaneously), and detection (AOM-1 on only). Note that the signal frequency injected into the AOM for Doppler cooling is red-shifted by about 5~MHz (the optical frequency shift is 10~MHz) from the resonant frequency to achieve the best cooling efficiency. The fluorescence of the ions during detection was collected by an objective lens with a numerical aperture of 0.4, resulting in a state preparation and measurement (SPAM) error of less than $7\times10^{-3}$.

The other beam diffracted from the AOM-2 is used to engineer the effective gain and damping processes in the targeted qubit system. The polarization of this beam is optimized to be perpendicular to the direction of the magnetic \textit{B} field, thus inducing only $\sigma_+$ and $\sigma_-$ transitions between the $^2S_{1/2}$ and $^2P_{1/2}$ levels (as shown in Fig.1~(c) in the main text or \figref{fig:Suppfig4} in the next section). The coupling strengths of the \textit{gain} beam ($\Omega_g$) and the \textit{damping} beam ($\Omega_d$) are determined by the sideband amplitudes generated by the 14.7~GHz and 2.11~GHz EOMs, respectively; therefore, they can be tuned independently by adjusting the signal amplitudes injected into each EOM. The carrier component after passing through the EOMs works as the \textit{repumping} beam. In the next section, we would derive the relationships between the values of $\Gamma_g$, $\Gamma_d$ and the values of $\Omega_g$, $\Omega_d$. The external driving signal for synchronization is generated by the frequency mixing and then amplified by a 10~Watt microwave amplifier. The frequency, the amplitude and the phase of the driving signal can be precisely controlled by tuning the DDS settings.

\section{Effective dissipated qubit system}

\begin{figure}[htbp!]
  \centering
  \includegraphics[scale = 1.]{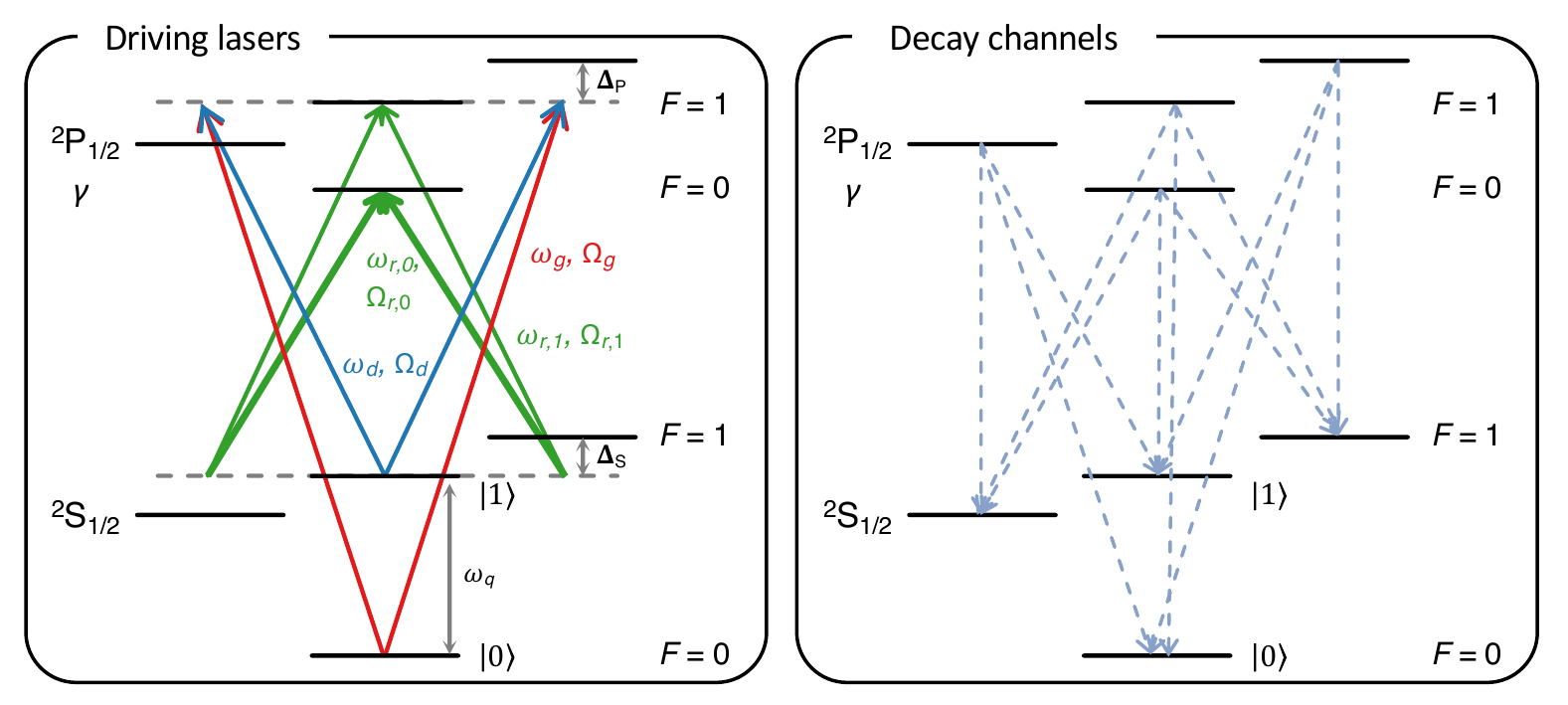}
  \caption{\label{fig:Suppfig4} Energy levels to engineer dissipated qubit system. Left: all the transitions driven by different laser beams with the optical frequency $\omega_{l}$ and coupling strength $\Omega_{l}$. Right: all possible decay channels from the excited P levels. Note that for each level there are three channels with equal decay probability.}
\end{figure}

Here, we introduce in theory that how we could obtain an effective dissipated qubit by engineering a multi-level system. Specifically, we utilize the eight-level system of $\yb$ to induce controllable damping rate $\Gamma_d$ and gain rate $\Gamma_g$ within the target qubit system, as shown in Fig.~1~(c) of the main text and \figref{fig:Suppfig4}.
The Hilbert space of the whole eight-level system can be divided into two subspaces, $\mathcal{H}_{t}$ and $\mathcal{H}_{a}$. The target Hilbert space $\mathcal{H}_{t}$ includes two levels of $\{ \sket{{0,0}} \equiv \ket{0}, \sket{{1,0}} \equiv \ket{1} \}$, while the remaining six levels $\{ \sket{{1,\pm1}}, \pket{{0,0}},\pket{{1,0}},\pket{{1,\pm1}} \}$ make up the auxiliary Hilbert space $\mathcal{H}_{a}$.
Here, the notation of $\ket{^{2S+1}L^{F,m_F}_{J}}$ represents the energy level $^{2S+1}L_{J}\ket{F,m_F}$. As shown in \figref{fig:Suppfig4}, we experimentally set $\Omega_{d},\Omega_{g},\Omega_{r,1}\ll\Omega_{r,0}\sim\gamma$, where $\Omega_{d,g,r}$ are the coupling strengths of the applied lasers and $\gamma$ is the spontaneous emission rate. This setting results in that the ion stays in target Hilbert space for almost all time, and then we could adiabatically eliminate all the auxiliary states and obtain an effective Lindblad master equation for the two-level open system~\cite{PRA2012effective}.

To derive the effective master equation, we follow the adiabatic elimination method introduced in Ref.~\cite{PhysRevA.85.032111}. For a general open system, we have the Lindblad master equation,
\begin{eqnarray}\label{mseq-0}
\frac{d\hat{\rho}}{dt}=-i[\hat{H},\hat{\rho}]+\sum_{k}\mathcal{D}\left[\hat{L}_{k}\right]\hat{\rho},
\end{eqnarray}
where $\mathcal{D}[\hat{\mathcal{A}}]\hat{\rho}=\hat{\mathcal{A}}\hat{\rho}\hat{\mathcal{A}}^{\dag}-\frac{1}{2}\{\hat{\mathcal{A}}^{\dag}\hat{\mathcal{A}},\hat{\rho}\}$ is the Lindblad superoperator, and $\hat{H}=\hat{H}_{t}+\hat{H}_{a}+\hat{V}_{t}+\hat{V}_{a}$. $\hat{H}_{t} ~(\hat{H}_{a})$ is the free Hamiltonian of the target (auxiliary) Hilbert space. $\hat{V}_{t}~(\hat{V}_{a})$ are the couplings with target (auxiliary) Hilbert space being the destination levels. Base on the perturbation theory and the adiabatic elimination method, \equref{mseq-0} can be reduced to an effective master equation of the target Hilbert space as below,
\begin{eqnarray}\label{mseq00}
\frac{d\hat{\rho}_{t}}{dt}=-i[\hat{H}_{\text{eff}},\hat{\rho}_{t}]+\sum_{k}\mathcal{D}\left[\hat{L}^{k}_{\text{eff}}\right]\hat{\rho}_{t},
\end{eqnarray}
where $\hat{\rho}_{t}$ represents the density matrix of the targed system, and the effective Hamiltonian $\hat{H}_{\text{eff}}$ reads,
\begin{eqnarray}\label{effHL1}
\hat{H}_{\text{eff}}&=&-\frac{1}{2}\left[\hat{V}_{t}(t)\sum_{l,n}(\hat{H}_{\text{NH}}^{(l,n)})^{-1}\hat{V}_{a}^{(l,n)}(t)+\text{H.c.}\right]+\hat{H}_{t}.
\end{eqnarray}
Moreover, the effective Lindblad operators in \equref{mseq00} can be expressed as,
\begin{eqnarray}\label{effHL2}
\hat{L}^{k}_{\text{eff}}&=&\hat{L}_{k}\sum_{l,n}(\hat{H}_{\text{NH}}^{(l,n)})^{-1}\hat{V}_{a}^{(l,n)}(t).
\end{eqnarray}
Here, $(\hat{H}_{\text{NH}}^{(l,n)})^{-1}=(\hat{H}_{\text{NH}}-E_{n}-\omega_{l})^{-1}$, and $\hat{V}_{a}^{(l,n)}$ is the coupling element for the field $l$ with the frequency of $\omega_l$ and the associated initial state is $|n\rangle$ with energy $E_{n}$. The non-Hermitian Hamiltonian $\hat{H}_{\text{NH}}$ in \equref{effHL1} and \equref{effHL2} reads as below,
\begin{eqnarray}\label{hnh}
\hat{H}_{\text{NH}}=\hat{H}_{a}-\frac{i}{2}\sum_{k}\hat{L}_{k}^{\dag}\hat{L}_{k}.
\end{eqnarray}

For our situation, the free Hamiltonian of target subspace can be written as,
\begin{eqnarray}
\hat{H}_{t}= \dfrac{\omega_q}{2}\hat\sigma_z,
\end{eqnarray}
and that of auxiliary subspace is,
\begin{eqnarray}\label{yba}
  \hat{H}_{a} =
    \sum_{m = \pm1} E_\mathrm{S}^{1,m_F} \sket{{1,m_F}} \sbra{{1,m_F}} +
    \sum_{F, m_F} E_\mathrm{P}^{F,m_F} \pket{{F,m_F}} \pbra{{F,m_F}}.
\end{eqnarray}
Here, $E_{L}^{F, m_F}$ is the eigenenergy of the energy level $\ket{ ^{2S+1}L^{F,m_F}_J}$, and the qubit energy gap $\omega_q$ is equal to $E_\mathrm{S}^{1,0}-E_\mathrm{S}^{0,0}$. The coupling operators induced by the driving lasers (shown in \figref{fig:Suppfig4}) can be summarized as,
\begin{eqnarray}\label{mseq01}
  \hat{V}_{t} &=&
    \dfrac{1}{2}\left(
        \Omega_{g}e^{i\omega_{g}t}\ket{0}
        + \Omega_{d}e^{i\omega_{d}t}\ket{1}
    \right)
    \left(
      \pbra{{1,-1}}+ \pbra{{1,1}}
    \right),\\
  \hat{V}_{a} &=& \hat{V}_{t}^\dagger +
    \left[
      \dfrac{1}{2}(\Omega_{r,0}e^{i\omega_{r,0}t}\pket{{0,0}}
      +\Omega_{r,1}e^{i\omega_{r,1}t}\pket{{1,0}})
      (\sbra{{1,-1}} + \sbra{{1,1}})+\text{H.c.}
    \right].
\end{eqnarray}
The spontanous emission from the excited P levels can be described as the the Lindblad operators as below,
\begin{eqnarray}\label{mseq02}
  \hat{L}^{(F,m_F)\rightarrow(F',m_{F'})} =
    \sqrt{\dfrac{\gamma}{3}}
    \sket{{F', m_{F'}}}\pbra{{F, m_{F}}}.
\end{eqnarray}
All the possible decay channels are listed in the right subfigure of \figref{fig:Suppfig4}. The coefficient of $\sqrt{\gamma/3}$ is because that each P level has three decay channels with equal probability. After inserting \equref{yba} and \equref{mseq02} into \equref{hnh}, we can obtain the following non-Hermitian Hamiltonian $\hat{H}_\mathrm{NH}$,
\begin{eqnarray}\label{mseq03}
  \hat{H}_{\text{NH}} =
  \hat{H}_{a} -
  \frac{i\gamma}{2}
  \sum_{F,m_F} \pket{{F,m_F}}\pbra{{F,m_F}}.
\end{eqnarray}
Now we can calculte the effective Hamiltonian of the targed qubit system and we find that,
\begin{eqnarray}\label{effHL3}
  \hat{H}_\mathrm{eff} & = &
  \left(
    -\dfrac{\Omega_g^2 \Delta_\mathrm{P}}
    {4\Delta_\mathrm{P}^2 + \gamma^2}
    -\dfrac{\Omega_g^2 (-\Delta_\mathrm{P})}
    {4\Delta_\mathrm{P}^2 + \gamma^2}
  \right) \ket{0}\bra{0}
  + \left(
    -\dfrac{\Omega_d^2 \Delta_\mathrm{P}}
    {4\Delta_\mathrm{P}^2 + \gamma^2}
    -\dfrac{\Omega_d^2 (-\Delta_\mathrm{P})}
    {4\Delta_\mathrm{P}^2 + \gamma^2}
  \right) \ket{1}\bra{1} \nonumber \\
  && -\dfrac{1}{2}
  \left(
    \ket{0}\bra{1}
    \dfrac{\Omega_g \Omega_d (\Delta_\mathrm{P} + (-\Delta_\mathrm{P}))}
    {4\Delta_\mathrm{P}^2 + \gamma^2}
    e^{i (\omega_g - \omega_d) t} + \mathrm{H. c.}
  \right) \nonumber \\
  && + \hat{H}_{t} \\
  & = & \hat{H}_{t}.
\end{eqnarray}
The first and second terms of the above effective Hamiltonian correspond to the A.C. Stark shift induced by the laser fields, while the third term represents the stimulated Raman transition induced by the beatnote of the laser beams. If the applied laser fields perfectly drive $\pket{{1,\pm 1}}$ levels with the symmetric detunings and equal strengths, both of these side-effects would be cancelled out. In our experimental demonstration, we found that the A.C. Stark shift is almost neglectable, while the Raman effect can be minimized by tuning $\omega_g - \omega_d$ a little bit away from $\omega_q$.

The engineered dissipation within the qubit system can be characterize by the effective Lindblad operators, which are written as,
\begin{equation}
  \hat{L}^{(1,\pm1)\rightarrow(F,0)}_\mathrm{eff} =
  \sqrt{\dfrac{\gamma}{3}}
  \dfrac{\Omega_g e^{-i \omega_g t}}
  {\pm2\Delta_\mathrm{P} - i\gamma}
  \ket{F}\bra{0}
  +
  \sqrt{\dfrac{\gamma}{3}}
  \dfrac{\Omega_d e^{-i \omega_d t}}
  {\pm2\Delta_\mathrm{P} - i\gamma}
  \ket{F}\bra{1}.
\end{equation}
Note that, the ion stays at the levels of $\pket{{1,\pm1}}$ would also have 1/3 probability to decay to the levels of $\sket{{1,\pm1}}$, respectively, and the strong repumping transitions from $\sket{{1,\pm1}}$ to $\pket{{1,0}}$ and $\pket{{1,1}}$ have the coupling strengths of $\Omega_{r,1}\ll\Omega_{r,0}\sim\gamma$. Therefore, we assume that the repumping process occurs almost instantaneously, and the effective Lindblad operators with the instantaneous repumping can be modified as,
\begin{equation}\label{l1}
  \hat{L}^{(1,\pm1)\rightarrow(1,0)}_\mathrm{eff, repump} =
  \sqrt{\dfrac{2\gamma}{3}}
  \dfrac{\Omega_g e^{-i \omega_g t}}
  {\pm2\Delta_\mathrm{P} - i\gamma}
  \ket{1}\bra{0}
  +
  \sqrt{\dfrac{2\gamma}{3}}
  \dfrac{\Omega_d e^{-i \omega_d t}}
  {\pm2\Delta_\mathrm{P} - i\gamma}
  \ket{1}\bra{1},
\end{equation}
\begin{equation}
  \hat{L}^{(1,\pm1)\rightarrow(0,0)}_\mathrm{eff, repump} =
  \sqrt{\dfrac{\gamma}{3}}
  \dfrac{\Omega_g e^{-i \omega_g t}}
  {\pm2\Delta_\mathrm{P} - i\gamma}
  \ket{0}\bra{0}
  +
  \sqrt{\dfrac{\gamma}{3}}
  \dfrac{\Omega_d e^{-i \omega_d t}}
  {\pm2\Delta_\mathrm{P} - i\gamma}
  \ket{0}\bra{1}.
\end{equation}
Here, the coefficients in \equref{l1} are modified from $\sqrt{\gamma/3}$ to $\sqrt{2\gamma/3}$, because we assume that the leakage to the levels of $\sket{{1,\pm1}}$ is almost repumped back to the $\ket{1}$ state under the situation of $\Omega_{r,0} \gg \Omega_{r,1}$. Then the effective Lindblad superoperators of the target Hilbert subspace turn out to be,
\begin{eqnarray}
  \mathcal{D}[\hat{L}^{(1,\pm1)\rightarrow(1,0)}_\mathrm{eff, repump}]\hat{\rho}_{t} & \approx &
  \dfrac{2\gamma}{3}
  \dfrac{\Omega_g^2}{4\Delta_\mathrm{P}^2 + \gamma^2}
  \mathcal{D}[\hat\sigma_+]\hat{\rho}_{t}
  +
  \dfrac{2\gamma}{3}
  \dfrac{\Omega_d^2}{4\Delta_\mathrm{P}^2 + \gamma^2}
  \mathcal{D}[\ket{1}\bra{1}]\hat{\rho}_{t},
  \label{effHL51}\\
  \mathcal{D}[\hat{L}^{(1,\pm1)\rightarrow(0,0)}_\mathrm{eff, repump}]\hat{\rho}_{t} & \approx &
  \dfrac{\gamma}{3}
  \dfrac{\Omega_g^2}{4\Delta_\mathrm{P}^2 + \gamma^2}
  \mathcal{D}[\ket{0}\bra{0}]\hat{\rho}_{t}
  +
  \dfrac{\gamma}{3}
  \dfrac{\Omega_d^2}{4\Delta_\mathrm{P}^2 + \gamma^2}
  \mathcal{D}[\hat\sigma_-]\hat{\rho}_{t}.
  \label{effHL52}
\end{eqnarray}
All the high-frequency oscillating terms are ignored, because in our experimental setup, the frequecy difference of $\omega_{g}-\omega_{d}$ is much great than the energy scale of the decay and coupling strengths. Combining \equref{effHL51} and \equref{effHL52}, we can finally obtain the effective dissipation on the qubit system as below,
\begin{eqnarray}\label{effHL6}
  \mathcal{L}_{0}\hat{\rho}_{t} & = &
  \sum_{ m = \pm1, F = 0,1}
  \mathcal{D}[\hat{L}^{(1,m)\rightarrow(F,0)}_\mathrm{eff, repump}]\hat{\rho}_{t} \nonumber \\
  & = &
  \frac{\Gamma_g}{2}\mathcal{D}[\hat\sigma_+]\hat{\rho}
  + \frac{\Gamma_d}{2}\mathcal{D}[\hat\sigma_-]\hat{\rho}
  + \frac{\Gamma_z}{2}\mathcal{D}[\hat\sigma_z]\hat{\rho},
\end{eqnarray}
where
\begin{eqnarray}
  \Gamma_g & = &
  \dfrac{8\gamma}{3}
  \dfrac{\Omega_g^2}{4\Delta_\mathrm{P}^2 + \gamma^2}, \\
  \Gamma_d & = &
  \dfrac{4\gamma}{3}
  \dfrac{\Omega_d^2}{4\Delta_\mathrm{P}^2 + \gamma^2}, \\
  \Gamma_z & = &
  \dfrac{\gamma}{3}
  \dfrac{2\Omega_d^2 + \Omega_g^2}{4\Delta_\mathrm{P}^2 + \gamma^2}. \\
\end{eqnarray}
Moreover, the effective Lindblad master equation can be written as,
\begin{eqnarray}\label{mseq0}
\frac{d\hat{\rho}_{t}}{dt}=-i[\hat{H}_{t},\hat{\rho}_{t}]+\mathcal{L}_{0}\hat{\rho}_{t}.
\end{eqnarray}
Note that, the values of $\Gamma_g$ and $\Gamma_d$ can be controlled respectively by adjusting the strengthes of $\Omega_g$ and $\Omega_d$, while $\Gamma_z$ is determined afterwards, which can not be independently tuned.

\section{Measurement of gain and damping rates}
\begin{figure}[htbp!]
  \centering
  \includegraphics[scale = 1.]{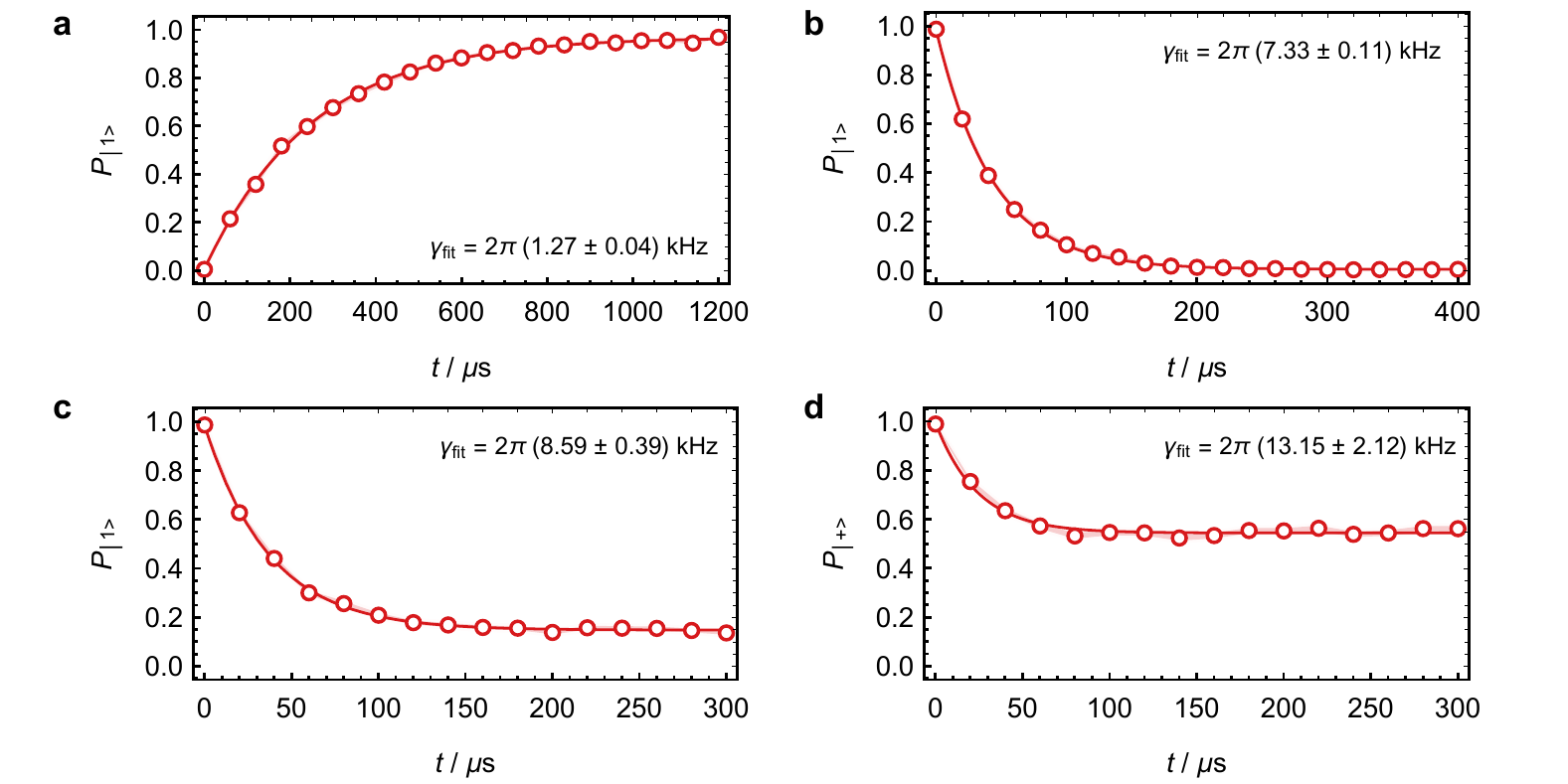}
  \caption{\label{fig:Suppfig2} Experimental results of measuring decay rates. \textbf{a} and \textbf{b} correspond to the pure gain and pure damping process, respectively. In \textbf{c} and \textbf{d}, both gain and damping process are engineered. In \textbf{c} we prepare the initial state to the $\ket{1}$ state and then measure it in the $\sigma_z$ basis, while in \textbf{d} we initialize the qubit state to $\ket{+} = (\ket{1} + \ket{0})/\sqrt{2}$ and then measure it in the $\sigma_x$ basis. Here, we utilize the fit function of $A e^{-\gamma t/2} + B$ to extract the decay rate $\gamma_\mathrm{fit}$ in each figure. The openmarkers and the shadow lines are the experimental results and the standard deviations respectively, and the solid red lines refer to the fitting results.}
\end{figure}

In our experiments, we independently evaluate the value of the gain rate $\Gamma_g$ (the damping rate $\Gamma_d$) by initialize the qubit state to the $\ket{0}$ ($\ket{1}$) state and then only engineering the gain (damping) process. The corresponding results are shown in \figref{fig:Suppfig2}~(a) and (b). Here, the gain rate $\Gamma_g$, the damping rate $\Gamma_d$ are $(2\pi)\times1.27(4)$~kHz and $(2\pi)\times7.33(11)$~kHz respectively according to the direct fitting results. In \figref{fig:Suppfig2}~(c) we simultaneously engineer the gain and the damping processes, and the measured decay rate of $(2\pi)\times8.59(39)$~kHz matches the value of $\Gamma_g + \Gamma_d$ well.

 To estimate the value of $\Gamma_z$, we initialize the qubit state to the eigenstate of $\sigma_x$ and then measure in the $\sigma_x$ basis as well, as shown in \figref{fig:Suppfig2}~(d). The decay rate here is equal to $2\Gamma_z + (\Gamma_g + \Gamma_d)/2$. Based on the values of $\Gamma_g$ and $\Gamma_d$ we have already obtained, we estimate the value of $\Gamma_z$ to be $(2\pi)\times4.42(106)$~kHz.

\section{Theory of Synchronizing Qubit with External Signal}
In this section, we summarize the theory of quantum synchronization of a two-level system (qubit) with an external drive signal. Most of the results have been given in Ref.~\cite{PRA2020two}. We show that, the additional dephasing term in \equref{effHL6} does not affect the ocurrence of synchronization. Moreover, as a complement to previous theoretical works, we also numerically simulate the driven dissipated qubit in the Schr\"{o}dinger picuture. In this case, the phenomenon of frequency synchronization can be captured more clearly.

\subsection{Representations of qubit state}
Before discussing the quantum synchronization of qubit, we first review the methods for representing the states of arbitrary qubit states. In addition to the conventional way of representing qubit states by density matrices, another common approach is to use the Bloch representation. In the Bloch representation, the density matrix of the qubit state is expanded in the basis of the Pauli matrices $\bm{\sigma} = \left\{ \sigma_x, \sigma_y, \sigma_z \right\}$, as follows,
\begin{eqnarray}
\rho=\frac{1}{2}(I+{\bf{m}}\cdot{\bm{\sigma}}),
\end{eqnarray}
The \textbf{m} here is known as the Bloch vector, and the relations between the vector components and the elements of the density matrix are shown as below,
\begin{eqnarray}
m_{x}&=&\rho_{01}+\rho_{10},\\
m_{y}&=&-i(\rho_{01}-\rho_{10}),\\
m_{z}&=&\rho_{11}-\rho_{00}=2\rho_{11}-1.
\end{eqnarray}
These show that the quantum coherence is determined by $m_{x,y}$ and the population distribution is encoded in $m_z$.

However, when we discuss the quantum synchronization of spin systems, it is more appropriate to use the Husimi-$Q$ representation. Here, utilizing a overcomplete set of spin-coherent states $\{|\theta,\phi\rangle, \theta\in[0,\pi],\phi\in(-\pi,\pi]\}$, arbitrary spin states can be mapped to quasi-probability distributions on the surface of the Bloch sphere.
Each spin-coherent state $|\theta,\phi\rangle$ is corresponding to one point on the Bloch sphere along the direction $\bf{n}_{\theta,\phi}=(\cos\phi\sin\theta,\sin\phi\sin\theta,\cos\theta)$, and for the single qubit system, this state can be generated by rotating the excited state as follows,
\begin{eqnarray}
|\theta,\phi\rangle=e^{-i \phi \sigma_z/2}e^{-i \theta \sigma_{y}/2}|1\rangle.
\end{eqnarray}
Similar to the coherent states of the quantized harmonic oscillator, the spin-coherent states are usually non-orthogonal to each other as well. The overlap of two spin-coherent states is given by,
\begin{eqnarray}
|\langle\theta',\phi'|\theta,\phi\rangle|^2=\frac{1+\bf{n}_{\theta',\phi'}\cdot\bf{n}_{\theta,\phi}}{2},
\end{eqnarray}
and the completeness relation reads as below,
\begin{eqnarray}
\frac{1}{2\pi}\int_{0}^{\pi}d\theta\sin\theta\int_{0}^{2\pi}d\phi |\theta,\phi\rangle\langle\theta,\phi|=1.
\end{eqnarray}

In the Husimi-$Q$ representation, we employ the $Q$-function, given by the following relation,
\begin{eqnarray}
Q(\theta,\phi)=\frac{1}{2\pi}\langle\theta,\phi|\rho|\theta,\phi\rangle,
\end{eqnarray}
to describe the quasi-probability distribution of arbitrary qubit state $\rho$ in the Bloch phase space. The Husimi-$Q$ representation allows us to visualize quantum spin state in terms of the most "classical" spin-coherent states, similar to the Wigner representation used for oscillator systems. Moreover, the $Q$-function can also be expressed in terms of the Bloch vector, given by the following relation,
\begin{eqnarray}
Q(\theta,\phi) &=& \frac{1}{4\pi} \left( 1 + \bf{m}\cdot\bf{n}_{\theta,\phi} \right) \\
&=&\frac{1}{4\pi}\left[1+(m_{x}\cos\phi+m_{y}\sin\phi)\sin\theta+m_{z}\cos\theta\right].
\end{eqnarray}

\subsection{Limit cycle}
As we have already mentioned in the main text, the limit cycle is one of the essential elements for a system to be able to synchronize. In order to maintain a self-sustained oscillation, a system that holds a valid limit cycle is always under dissipation.
For the single qubit system, we consider it in a generalized open environment in which the dynamics of the system can be expressed as the following Lindblad equation, 
\begin{eqnarray}\label{mseq1}
\frac{\mathrm{d}\hat{\rho}}{dt}
  =
  - i \left[ \hat{H}_q, \hat{\rho} \right]
  + \frac{\Gamma_g}{2}\mathcal{D}[\hat\sigma_+]\hat{\rho}
  + \frac{\Gamma_d}{2}\mathcal{D}[\hat\sigma_-]\hat{\rho}
  + \frac{\Gamma_z}{2}\mathcal{D}[\hat\sigma_z]\hat{\rho},
\end{eqnarray}
where $\hat{H}_{q} = \omega_q \hat{\sigma}_z /2$ is the free Hamiltonian of a two-level system with the energy gap of $\omega_q$, and the latter three terms correspond to the gain, damping and dephasing processes, respectively. As one of the standard approaches for studying two-level systems, we can translate above master equation into the Bloch representation, leading to an equation array of,
\begin{eqnarray}
\dot{m}_{x} &=& -\dfrac{\Gamma_t m_{x}}{4} - \omega_q m_y,\\
\dot{m}_{y} &=& -\dfrac{\Gamma_t m_{y}}{4} + \omega_q m_x,\\
\dot{m}_{z} &=& \dfrac{1}{2}\left[\Gamma_{g}(1-m_z)-\Gamma_{d}(1+m_z) \right],
\end{eqnarray}
where $\Gamma_t = \Gamma_g + \Gamma_d + 4\Gamma_z$. Obviously that above equations has a stationary solution of,
\begin{eqnarray}
  \mathbf{m}_\mathrm{LC} =\left\{ 0, 0, \frac{\Gamma_{g}-\Gamma_{d}}{\Gamma_{g}+\Gamma_{d}} \right\},
\end{eqnarray}
or
\begin{equation}\label{lcdensitym}
  \rho_\mathrm{LC} =
    \dfrac{\Gamma_d}{{\Gamma_g + \Gamma_d}} \ket{0}\bra{0} +
    \dfrac{\Gamma_g}{{\Gamma_g + \Gamma_d}} \ket{1}\bra{1},
\end{equation}
if we translate the Bloch vector back to the density matrix form. Note that, there is seemingly no free phase if we simply treat the above stationary state in the basis of $\{ \ket{0}, \ket{1} \}$. It is why that in Ref. \cite{PRL2018}, they stated that there is no limit cycle for a single qubit and therefore the qubit can not be synchronized.

However, the conclusion turns out to be different if we choose the overcomplete set of spin-coherent states as the basis to represent the above stationary state. Here, we rewrite Eq. (\ref{lcdensitym}) by utilizing the spin-coherent states and the result is given by,
\begin{equation}\label{lc}
  \rho_\mathrm{LC} = \dfrac{1}{2\pi} \int_{0}^{2\pi} d\phi \ket{\theta_0, \phi}\bra{\theta_0, \phi},
\end{equation}
where
\begin{equation}
  \theta_0 = \arccos
    \left(
      \dfrac{\Gamma_g - \Gamma_d}{\Gamma_g + \Gamma_d}
    \right).
\end{equation}
The stationary state can be treat as an ensemble of pure states with the same energy, and all these pure states rotate along the $z$-axis at an instrinct frequency of $\omega_q$ according to the relation of, 
\begin{eqnarray}
  \hat{U}(t)|\theta, \phi\rangle = e^{-i \hat{H}_q t }|\theta,\phi\rangle = |\theta,\phi+\omega_{q}t\rangle.
\end{eqnarray}
Because the spin-coherent states with a certain energy and arbitrary phase $\phi$ contribute equally here, it is proper to treat $\phi$ as the free phase. Therefore, the stationary state resulted from the dynamics of \equref{mseq1} can be definitely interpreted as a valid limit cycle. Moreover, in the Husimi-$Q$ representation, we can also express the limit cycle state as follows,
\begin{eqnarray}
Q(\theta,\phi)&=&\frac{1}{4\pi}\left(1+m_{\text{LC}, z}\cos\theta\right),
\end{eqnarray}
which shows a uniform quasiprobability distribution along the longitude axis.

\subsection{Characterizing synchronization in the rotating frame}

After we establish a valid limit cycle in the dissipated qubit system, we apply a external sync signal with a strength $\epsilon$, frequency $\omega$ ($\omega \sim \omega_q$) and initial phase $\varphi$ to synchronize it. This signal induces a coherent transition within the qubit which reads as
\begin{eqnarray}
\hat{H}_\mathrm{sync} = \epsilon\hat{\sigma}_x\cos(\omega t + \varphi),
\end{eqnarray}
A standard way to deal with the above time-dependent interaction is choosing the rotating frame with respect to the frequency of the driving signal, denoted as $\hat{U}_\mathrm{sync} = e^{- i \omega \hat{\sigma}_z t /2}$. We can obtain the dynamics of the qubit in this rotating frame as follows,
\begin{eqnarray}
\hat{H}_\mathrm{sys} = \hat{U}_\mathrm{sync}^\dagger (\hat{H}_q + \hat{H}_\mathrm{sync} - \dfrac{\omega}{2} \hat{\sigma}_z) \hat{U}_\mathrm{sync} \approx \dfrac{\Delta}{2}\hat{\sigma}_z + \dfrac{\epsilon}{2} \hat{\sigma}_\varphi,
\end{eqnarray}
since the rotating wave approximation is valid under our experimental parameters ($\omega_q + \omega \gg |\omega_q - \omega|$). Here $\Delta = \omega_q - \omega$ is the detuning between the sync frequency and the qubit resonant frequency, and $\sigma_\varphi = \sigma_x \cos\varphi  + \sigma_y \sin\varphi$. After including the gain, damping and dephasing processes, the system dynamics thus reads as follows,
\begin{eqnarray}\label{mseq2}
\frac{\mathrm{d}\hat{\rho}_s}{dt}
  = -i[\hat{H}_\mathrm{sys},\hat{\rho}_s] + \frac{\Gamma_g}{2}\mathcal{D}[\hat{\sigma}_+]\hat{\rho}_s + \frac{\Gamma_d}{2}\mathcal{D}[\hat{\sigma}_-]\hat{\rho}_s + \frac{\Gamma_z}{2}\mathcal{D}[\hat{\sigma}_z]\hat{\rho}_s,
\end{eqnarray}
where $\hat{\rho}_s = \hat{U}_\mathrm{sync}^\dagger\hat{\rho} \hat{U}_\mathrm{sync}$. The corresponding equations of the Bloch vector can be written as,
\begin{eqnarray}
\dot{m}_{x}&=& -\dfrac{\Gamma_t m_{x}}{4} - \Delta m_{y} + \epsilon \cos\varphi m_{z},\\
\dot{m}_{y}&=&- \dfrac{\Gamma_t m_{y}}{4} + \Delta m_{x} + \epsilon \sin\varphi m_{z},\\
\dot{m}_{z}&=& \dfrac{1}{2}\left[\Gamma_{g}(1-m_z)-\Gamma_{d}(1+m_z)\right] - \epsilon\cos\varphi m_{x} - \epsilon\sin\varphi m_y,
\end{eqnarray}
and thus the stationary solutions turn out to be,
\begin{eqnarray}
m_{x} &=&
  \frac
    {4\epsilon(\Gamma_{g}-\Gamma_{d})}
    {(16\Delta^{2}+\Gamma_t^{2})(\Gamma_{g}+\Gamma_{d})+8\Gamma_t\epsilon^{2}}(4\Delta\cos\varphi + \Gamma_t \sin\varphi),\label{sm1}\\
m_{y} &=&
  \frac
    {4\epsilon(\Gamma_{g}-\Gamma_{d})}
    {(16\Delta^{2}+\Gamma_t^{2})(\Gamma_{g}+\Gamma_{d})+8\Gamma_t\epsilon^{2}}(4\Delta\sin\varphi - \Gamma_t \cos\varphi),\label{sm2}\\
m_{z} &=&
  \frac
    {(16\Delta^{2}+\Gamma_t^{2})(\Gamma_{g}-\Gamma_{d})}
    {(16\Delta^{2}+\Gamma_t^{2})(\Gamma_{g}+\Gamma_{d})+8\Gamma_t\epsilon^{2}}\label{sm3}.
\end{eqnarray}

To characterize the performance of synchronization after applying the external driving signal and then reaching to the steady state, we utilize the \textit{synchronization measurement} defined as,
\begin{eqnarray}
S(\phi)&=&\int_{0}^{\pi}d\theta Q(\theta,\phi)  \sin\theta-\frac{1}{2\pi},
\end{eqnarray}
which also can be rewritten into the Bloch representation as below,
\begin{equation}\label{sfunction}
  S(\phi)=\frac{1}{8}(m_{x}\cos\phi+m_{y}\sin\phi).
\end{equation}
Obviously, the synchronization measurement gives rise to $S(\phi)=0$ for the unsynchronized limit cycle state. While the qubit state is synchronized, the $S$-function would show a non-uniform distribution over $\phi$.

For the qubit with the external driving signal, we can analytically obtain the \textit{S}-function by inserting Eqs. (\ref{sm2}-\ref{sm3}) into Eq. (\ref{sfunction}) and the result turns out to be,
\begin{equation}\label{sfuncsync}
  S(\phi) = \dfrac{\mathcal{C}}{2} \sin(\varphi + \varphi_d - \phi),
\end{equation}
where the contrast $\mathcal{C}$ reads as,
\begin{equation}
  \mathcal{C} = \dfrac{\epsilon |\Gamma_g - \Gamma_d|}
  {
    (16\Delta^2 + \Gamma_t^2)(\Gamma_g + \Gamma_d) + 8\Gamma_t\epsilon^2}
  \sqrt{16\Delta^2 + \Gamma_t^2},
\end{equation}
and the phase shift $\varphi_d$ is,
\begin{equation}
  \varphi_d = \arctan\dfrac{4\Delta}{\Gamma_t}.
\end{equation}
From Eq. (\ref{sfuncsync}), we can straightforwardly obtain the synchronized phase,
\begin{equation}
  \phi_s =\left\{
  \begin{aligned}
    \varphi + \varphi_d - \dfrac{\pi}{2}, & ~~~\Gamma_g > \Gamma_d \\
    \varphi + \varphi_d + \dfrac{\pi}{2}, &~~~ \Gamma_g < \Gamma_d
  \end{aligned}
  \right..
\end{equation}
In our experimental demonstration we set $\Gamma_g < \Gamma_d$ and the phase of the external driving field to be $\varphi = \pi/2$. Therefore, for the resonant situation ($\Delta = 0$), the synchronized phase turns out to be $\pi$.

Note that, if the gain and damping rates are equal, Eqs. (\ref{sm1}-\ref{sm3}) turn out to be zero no matter what kinds of the driving signal is applied to the qubit system. Therefore, in this situation, we claim that the qubit is not synchronizable.

\subsection{Characterizing synchronization in the Schr\"{o}dinger picture}

In the above discussion, we have shown how to characterize the synchronization in a rotating frame, where the qubit state evolves into a time-independent steady state after enough long time. In the rotating frame, we can detect the feature of phase-locking, which is sufficient to account for the validity of synchronization. However, the fact of frequency synchronization is not straightforward enough to observe in the rotating frame. In principle, we can obtain the actual oscillating frequency if we can measure the qubit coherence directly in the non-rotating lab frame (or the Schr\"{o}dinger picture). However, it is hardly feasible in experiments since the intrinsic frequency of the qubit is too high ($\omega_q \sim 2\pi\times12.6$~GHz in our case).

Here, as a supplementary, we numerically simulate the dynamics of the quantum synchronization in the Schr\"{o}dinger picture by directly solving the following master equation,
\begin{eqnarray}\label{scheq}
  \frac{\mathrm{d}\hat{\rho}}{dt}
    =
    - i \left[ \dfrac{\omega_q}{2} \hat{\sigma}_z + \epsilon\hat{\sigma}_x\cos(\omega t + \varphi), \hat{\rho} \right]
    + \frac{\Gamma_g}{2}\mathcal{D}[\hat\sigma_+]\hat{\rho}
    + \frac{\Gamma_d}{2}\mathcal{D}[\hat\sigma_-]\hat{\rho}
    + \frac{\Gamma_z}{2}\mathcal{D}[\hat\sigma_z]\hat{\rho}.
  \end{eqnarray}
The feature of the frequency synchronization is clearly revealed by the simulation resutls.

In our numerical simulation, we set most of the parameters in \equref{scheq} to be consistent with the experimental settings,
\begin{eqnarray}
  \epsilon &=& 2\pi \times 2.37 ~\mathrm{kHz}, \\
  \varphi &=& \pi/2, \\
  \Gamma_g &=& 2\pi \times 1.27 ~\mathrm{kHz}, \\
  \Gamma_d &=& 2\pi \times 7.33 ~\mathrm{kHz}, \\
  \Gamma_z &=& 2\pi \times 4.42 ~\mathrm{kHz},
\end{eqnarray}
and sync frequency $\omega$ is a tunable parameter in the simulation. Meanwhile, to reduce the time overhead of the numerical simulation, without the loss of the generality, we assume the value of $\omega_q$ to be $2\pi\times 10$~MHz instead of the actual value of $2\pi\times 12.6$~GHz in the experiment. But the rotating wave approximation remains valid if we choose $\omega$ to make $|\omega_q - \omega|$ on the order of $\Gamma_g$; therefore, we can still use \equref{mseq2} to describe the dynamics of the qubit in the rotating frame.

\begin{figure}[htbp!]
  \centering
  \includegraphics[scale = 1.]{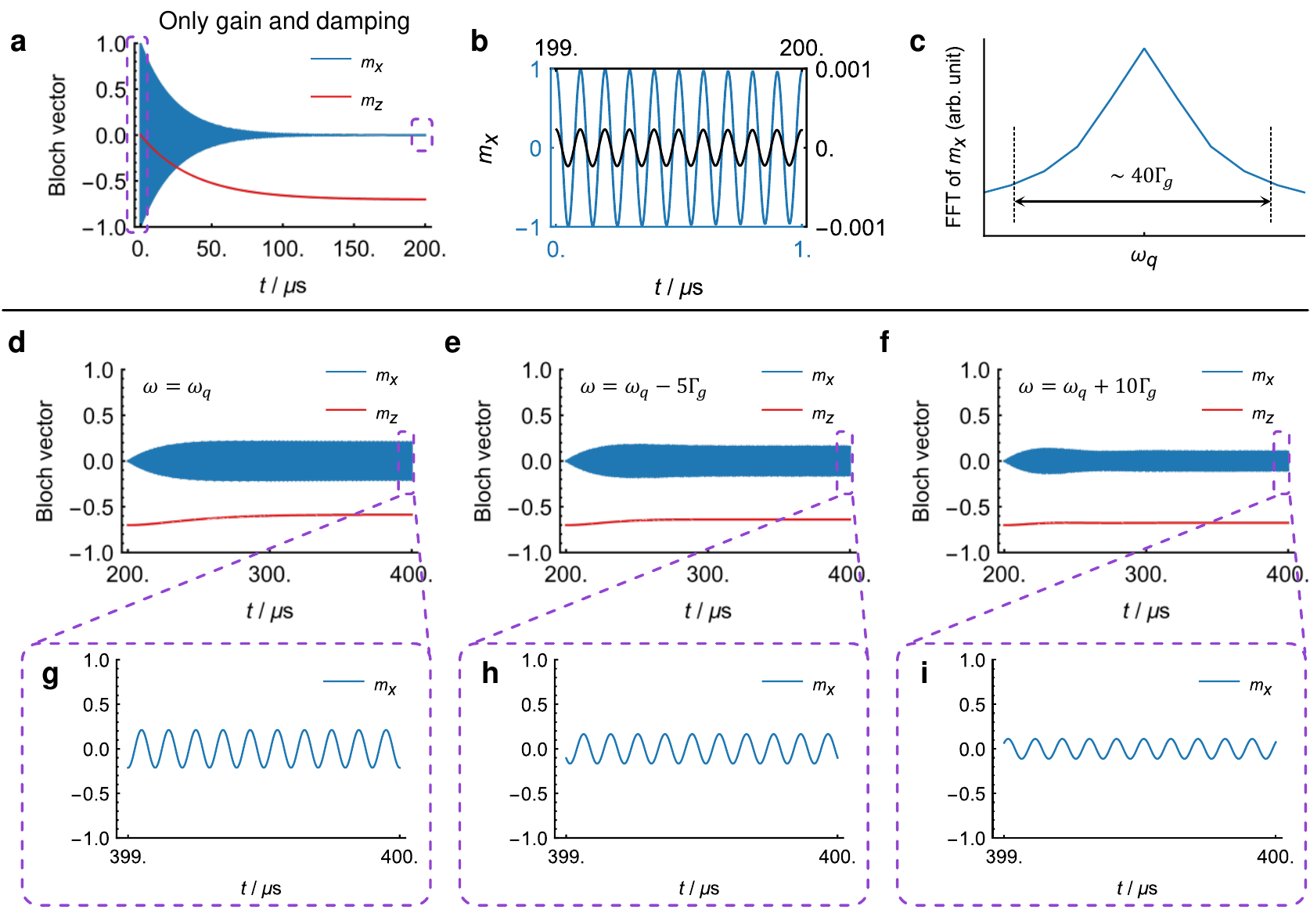}
  \caption{\label{fig:Suppfig5} 
    Quantum synchronization in the Schrodinger picture.
    \textbf{a}. Time evolution of the Bloch vector under only dissipations. The initial state is set to $(\ket{0}+\ket{1})/\sqrt{2}$.
    \textbf{b}. We zoom in on the time evolution of the Bloch vector shown in \textbf{a}, specifically focusing on the oscillations in the intervals (0., ~1.)~$\mu$s and (199.,~200.)~$\mu$s. The amplitude of the oscillation decays as expected while the oscillating frequency remains unchanged.
    \textbf{c}. We apply the fast Fourier transformation to the simulation data shown in \textbf{a}, with a sampling interval of 0.002~$\mu$s. The transformed spectral data show a distinct peak at the qubit's intrinsic frequency $\omega_q$.
    \textbf{d - f}. After the relaxation of the qubit to the limit cycle, we simulate the evolution of the qubit by adding external signals with different parameters. We set the values of the sync frequency to be $\omega_q$, $\omega_q - 5\Gamma_g$, $\omega_q + 10\Gamma_g$, corresponding to the results in \textbf{d} to \textbf{f}, respectively. For all cases, we can observe the recovery of the oscillations of the qubit coherence ($m_x$) while the energy of the qubit changes only slightly ($m_z$).
    \textbf{g - i}. We zoom in on the time evolution from 399.~$\mu$s to 400.~$\mu$s shown in \textbf{d} to \textbf{f}. The oscillations of the qubit coherence become stable, and the oscillation amplitude decays as the driving frequency of the external signal is tuned away from the qubit frequency.
    }
\end{figure}

\begin{figure}[htbp!]
  \centering
  \includegraphics[scale = 1.]{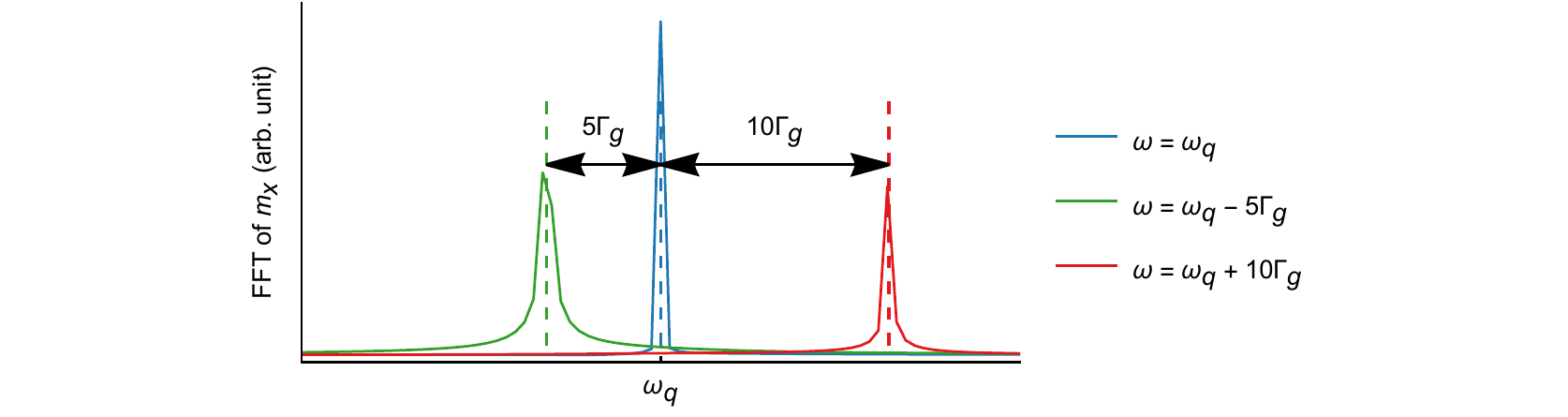}
  \caption{\label{fig:Suppfig6} Spectral analysis on the synchronized qubit. Based on the simulation data shown in \figref{fig:Suppfig5}~\textbf{d - f}, we apply the fast Fourier transformation to the data in the interval (300., 2000.)~$\mu$s (400~$\mu$s to 2000.~$\mu$s not shown in \figref{fig:Suppfig5}), with a sampling interval of 0.002~$\mu$s. It is clear that the qubit oscillates at the frequency of the external signal rather than its intrinsic frequency, indicating that the qubit is synchronized with the external signal.
  }
\end{figure}

Following the sequence we utilized in the experimental demonstration, we also divide the evolution into two stages in the numerical simulation, and the results are summarized in \figref{fig:Suppfig5}. During the first stage (lasting 200~$\mu$s), the qubit only suffers from the dissipations while the external sync signal is absent, allowing the qubit to relax to the limit cycle, as shown in \figref{fig:Suppfig5}~(a). Here, in order to illustrate the oscillation of the qubit coherence $m_{x}$, the initial state of the qubit is set to $(\ket{0}+\ket{1})/\sqrt{2}$ instead of the $\ket{1}$ used in the experiments. Due to the fast oscillation of the qubit coherence, it is hard to obtain details from \figref{fig:Suppfig5}~(a), but the oscillating profile exponentially decays as expected. In \figref{fig:Suppfig5}~(b), we further show the details of the oscillation by zooming in on the first and the last 1~$\mu$s evolution. It is clear that they occur at the same frequency. Additionally, we apply fast Fourier transformation to the $m_x$ data shown in \figref{fig:Suppfig5}~(a), which reveals a single peak at the frequency of $\omega_q$. This indicates that when external driving is absent, the qubit oscillates at its intrinsic frequency.

From 200~$\mu$s in our numerical simulation, an external driving signal is added. Here, we choose three values of the sync frequency, $\omega = \omega_q,~\omega_q-5\Gamma_g,~\omega_q + 10\Gamma_g$, to observe the respone of the qubit dynamics. The results are summarized in \figref{fig:Suppfig5} (d-f). From all the cases, we can clearly observe the recovery of the oscillation of the qubit coherence, corresponding to the finite values of $m_x^\mathrm{rot}$ and $m_y^\mathrm{rot}$ we observe in the rotating frame. Here, we use the superscript "$\mathrm{rot}$" to distinguish the coherence obtained in the rotating frame (with $\mathrm{rot}$) and in the lab frame (without $\mathrm{rot}$). We can find the relation that the oscillating amplitude of $m_x$ is equal to $\sqrt{ (m_x^\mathrm{rot})^2 + (m_y^\mathrm{rot})^2 }$.

\begin{figure}[htbp!]
  \centering
  \includegraphics[scale = 0.8]{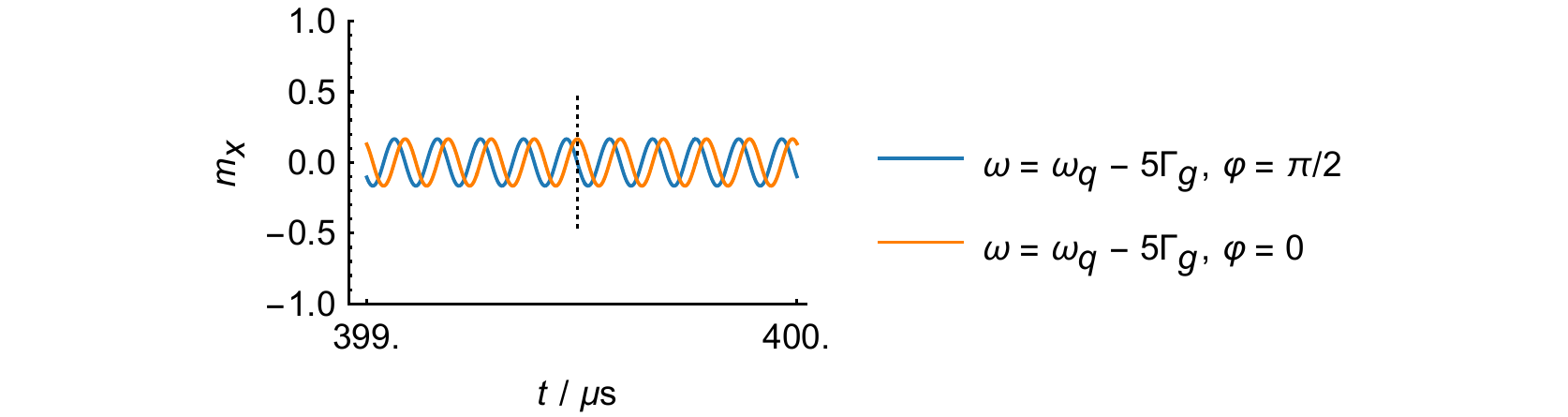}
  \caption{\label{fig:Suppfig6-2} Phase shift in qubit oscillations. Under the sync frequency of $\omega_q - 5\Gamma_g$, we shift the initial phase of the sync signal from $\varphi = 0$ to $\varphi = \pi/2$, and a $\pi/2$ phase shift is observed in the qubit oscillations as well.
  }
\end{figure}

Of course, we are particularly interested in the oscillation frequency. In \figref{fig:Suppfig5} (g-i), we zoom in on the last microsecond oscillations that are shown in \figref{fig:Suppfig5} (d-f), respectively. To quantify the frequency of the recovered oscillations, we perform the fast Fourier transform on the steady oscillation data from 300~$\mu$s to 2000~$\mu$s (the data points from 400~$\mu$s to 2000~$\mu$s are not shown in \figref{fig:Suppfig5} \textbf{d} to \textbf{f}). The corresponding spectral results are shown in \figref{fig:Suppfig6}. It is obvious to see three peaks at the frequencies of $\omega_q$, $\omega_q - 5\Gamma_g$ and $\omega_q + 10\Gamma_g$, which are well consistent with the corresponding sync frequencies of the driving signals. This evidences strongly confirms that the qubit is synchronized with the external driving signal and oscillates at the same frequency. Moreover, if we alter the starting phase of the sync signal, this phase shift can also be seen in the oscillation of the qubit coherence, as in \figref{fig:Suppfig6-2}.

\section{Experimental construction of \textit{Q}- and \textit{S}-function}
Here, we show how to obtain \textit{Q}-function and \textit{S}-function in experiments. It is straightforward to measure the above two functions if the sync signal is resonant to the qubit frequency, which means $\hat{U}_q = \hat{U}_s$. By applying single-qubit rotations on the desired qubit state $\hat{\rho}$ and then measuring the state probability on the $\ket{1}$ state, we can directly calculate the value of $Q(\theta,\phi)$ based on the relation of,
\begin{eqnarray}\label{equ:qfunc}
 \nonumber % Remove numbering (before each equation)
  Q(\theta,\phi) &=& \dfrac{1}{2\pi} \bra{\theta,\phi}\hat\rho\ket{\theta,\phi} \\
  &=& \dfrac{1}{2\pi} \mathrm{Tr}\left[  e^{-\mathrm{i}\theta \hat\sigma_{\phi-\pi/2}/2} \hat\rho e^{\mathrm{i}\theta \hat\sigma_{\phi-\pi/2}/2} (\ket{1}\bra{1}) \right],
\end{eqnarray}
where $\hat\sigma_{\phi} = \hat\sigma_x \cos\phi + \hat\sigma_y \sin\phi$. And then the corresponding \textit{S}-function can be constructed through the discrete integration of the $\left\{ Q(\theta_i, \phi_j) \right\}$ obtained in the experiment. Above methods are applied to obtain the \textit{Q}-functions and the \textit{S}-functions in Fig.~2~(b-g), and \textit{Q}-functions in the inset figures of Fig.~4~(a).

However, above methods to obtain the \textit{S}-function is in-efficient. Instead, we can just measure the Bloch vector $\mathbf{m}$ of the qubit state $\hat\rho$ and then use the relation of,
\begin{equation}\label{equ:sfunc}
  S(\phi) = \dfrac{1}{8} \left( m_x \cos\phi + m_y \sin\phi \right),
\end{equation}
to construct the \textit{S}-function. In the resonant driving case ($\hat{U}_q = \hat{U}_s$), we can obtain the components of the Bloch vecor by applying the analysis pulses in the set of
\begin{equation}\label{equ:Mset1}
  \left\{ e^{i \Omega_\mathrm{MW}\tau_\pi \sigma_y / 4}, e^{ -i \Omega_\mathrm{MW}\tau_\pi \sigma_x / 4}, I \right\},
\end{equation}
to the qubit and then measure the probability of projecting to the $\ket{1}$ state. Here $\Omega_\mathrm{MW},\tau_\pi$ are the coupling strength and the duration of the analysis pulse, respectively and satisfy the relation of $\Omega_\mathrm{MW}\tau_\pi = \pi$. The measurement results of each analysis pulse correspond to the value of $(m_x + 1)/2$, $(m_y + 1)/2$ and $(m_z + 1)/2$, respectively. In the experiments, we set $\Omega_{\mathrm{MW}}$ and $\tau_\pi$ to be $(2\pi)\times 32.0(1)$~kHz and $15.6~\mu\mathrm{s}$, respectively. However, when the sync signal is detuned away from the qubit resonant frequency, the above method fails because the synchronized state and the analysis pulses are not in the same rotating frame $(\hat{U}_q \neq \hat{U}_s)$. In this case, we use a set of non-orthogonal analysis pulses,
\begin{equation}\label{equ:Mset2}
  \left\{
    e^{- i (\Delta \sigma_z + \Omega_\mathrm{MW} \sigma_x)\tau_\pi /4},
    e^{- i (\Delta \sigma_z + \Omega_\mathrm{MW} \sigma_x)\tau_\pi /2},
    e^{- i (\Delta \sigma_z + \Omega_\mathrm{MW} \sigma_y)\tau_\pi /4},
    e^{- i (\Delta \sigma_z + \Omega_\mathrm{MW} \sigma_y)\tau_\pi /2},
    I
  \right\}
\end{equation}
depending on the detuning $\Delta$. The Bloch vector can be estimated by using Maximum likelihood estimation on the measurement results. This method is used to obtain the results in Fig.~3.

\section{Quantum synchronization under different sync strengthes}

\begin{figure}[htbp!]
\centering
\includegraphics[scale = 1.]{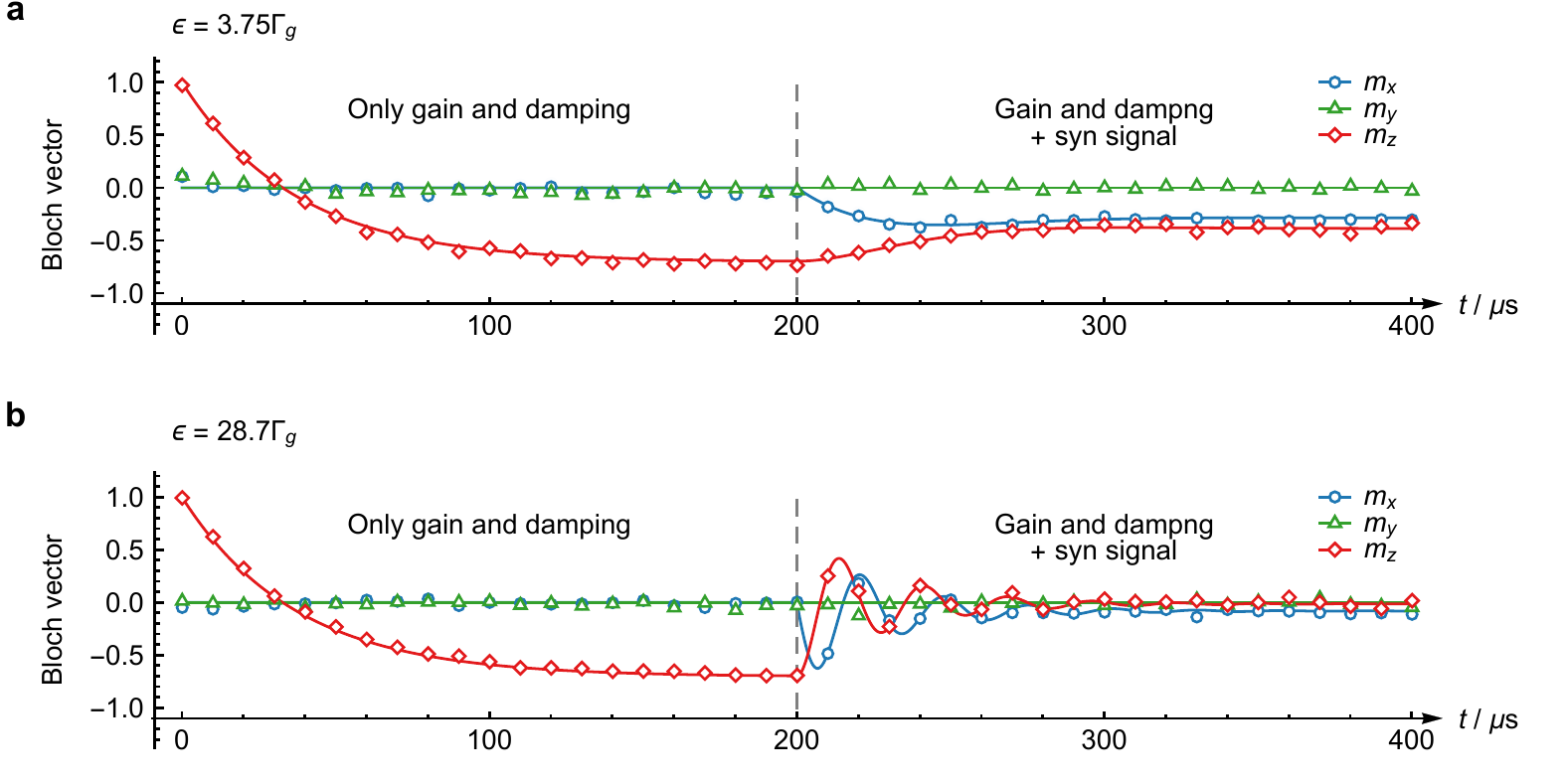}
\caption{\label{fig:Suppfig3} Time evolution of the Bloch vector under different sync strengths. The results in \textbf{a} and \textbf{b} correspond to the sync signals with the strengths of $3.75\Gamma_g$ and $28.7\Gamma_g$, respectively. The open markers and the solid lines represent the experiment and simulation results for each Bloch vector component.}
\end{figure}

In Fig.~4 in the main text, we show the experimental results about the time evolution of the Bloch vector component $m_z$ under different sync strengths. Here, we supplement the time evolution for all other components, as shown in \figref{fig:Suppfig3}.

\end{document}